\documentclass[a4paper,11pt]{article}
\pdfoutput=1 

\usepackage{jcappub} 

\usepackage[T1]{fontenc} 
\usepackage{subfigure}
\usepackage{multirow}
\usepackage{color}
\usepackage[]{hyperref}
\usepackage{comment}
\usepackage{here}
\usepackage{cancel}
\usepackage{amssymb}
\usepackage[mathscr]{euscript}
\usepackage{physics}
\usepackage{ulem}  
\usepackage{lineno}
\usepackage{amsmath}

\title{
 Probing neutrino decay scenarios by using the Earth matter effects on supernova neutrinos
}

\author[a]{Edwin A. Delgado,}
\author[a,b]{Hiroshi Nunokawa,}
\author[a]{and Alexander A. Quiroga}

\affiliation[a]{Departamento de F\'isica, Pontif\'icia Universidade Cat\'olica do Rio de Janeiro, C.P. 38071, 22452-970, Rio de Janeiro, Brazil}
\affiliation[b]{IJCLab, Universit\'e Paris-Saclay, CNRS/IN2P3, 91405 Orsay, France}

\emailAdd{a.delgado4d@aluno.puc-rio.br}
\emailAdd{nunokawa@puc-rio.br}
\emailAdd{alarquis@puc-rio.br}

\abstract{
The observation of Earth matter effects in the spectrum of neutrinos coming from a next galactic core-collapse supernova (CCSN) could, in principle, reveal if neutrino  mass ordering is normal or inverted. One of the possible ways to identify the mass ordering is through the observation of the modulations that appear in the spectrum when  neutrinos travel through the Earth  before they arrive at the detector. These features in the neutrino spectrum depend on two factors, the average  neutrino energies, and the difference between the  primary neutrino fluxes of electron and other flavors produced inside the supernova.  However,   recent studies indicate that the Earth matter effect  for CCSN neutrinos is expected to be rather small and difficult to be observed by currently operating or planned neutrino detectors mainly  because of the similarity of  average energies and fluxes between electron and other flavors of neutrinos, unless the distance to CCSN is significantly smaller than the typically expected one, $\sim 10$ kpc.  Here, we  are looking towards the possibility if the non-standard neutrino  properties such as decay of neutrinos can enhance the Earth matter effect.  In this work  we show that invisible neutrino decay can potentially enhance significantly the Earth matter effect for both $\nu_e$ and $\bar{\nu}_e$ channels at the same time for both mass orderings,  even if the neutrino spectra between electron and other  flavors of neutrinos are very similar,  which is  a different feature not expected for CCSN neutrinos with standard oscillation without the decay effect.
}

\begin{document}
\maketitle
\flushbottom


\section{Introduction}
\label{sec1}

Despite that so far only a small number of events of neutrinos coming from SN1987A were detected in February 1987 by Kamiokande~\cite{Hirata:1987hu}, IMB~\cite{Bionta:1987qt} and BAKSAN~\cite{Alekseev:1988gp} detectors, core collapse supernovae (CCSNe) are considered as one of the most important and interesting astrophysical neutrino sources, see e.g.~\cite{Scholberg:2012id,Mirizzi:2015eza,Horiuchi:2017sku} for reviews. Compared to the epoch when SN1987A neutrinos were observed, we have currently larger number and better capability of neutrino detectors in operation allowing us to learn much more about physics of CCSN as well as some unknown neutrino properties, once CCSN neutrinos are observed. In addition, by taking into account several new neutrino detectors which are currently in construction and will be in operation within this decade, such as JUNO~\cite{An:2015jdp,JUNO:2021vlw}, Hyper-Kamiokande~\cite{Abe:2018uyc}  and DUNE~\cite{Abi:2020evt} we have very good prospect in making a significant progress in improving our knowledge on CCSN neutrinos as well as physics related to CCSN explosion if they will be detected by these detectors in the near future. In this work we consider the possibility of decay of neutrinos, much faster than the rate expected for massive neutrinos possessing only standard electroweak interactions (but slow enough not to be inconsistent with observed neutrino data in terrestrial experiments) and its impact for the observation of CCSN neutrinos paying particular attention to the Earth matter effect.  Due to the much longer baseline, typically $\sim O(10)$ kpc, CCSN neutrinos are expected to be much more sensitive to the decay effect than the terrestrial neutrino experiments, see e.g.  \cite{Lindner:2001fx,Maltoni:2008jr,  Abrahao:2015rba,Choubey:2017dyu,Coloma:2017zpg,Gago:2017zzy,Choubey:2018cfz,Ascencio-Sosa:2018lbk,deSalas:2018kri,Porto-Silva:2020gma,Tabrizi:2020vmo}. For previous works which discussed the impact of decay effects for CCSN neutrinos, see e.g. refs.~\cite{Frieman:1987as,Chupp:1989kx,Beacom:2002cb,Fogli:2004gy,Ando:2003ie,Ando:2004qe,deGouvea:2019goq}.  For the decay impact on other astrophysical neutrinos, see e.g.~\cite{Lindner:2001th,Bustamante:2016ciw,Denton:2018aml}.

The Earth matter effect for CCSN neutrinos has been extensively  studied by many authors as it can be used to determine neutrino mass  ordering, see e.g.~\cite{Dighe:1999bi,Lunardini:2001pb,Dighe:2003jg,Mirizzi:2006xx,Borriello:2012zc,Machado:2012ee,Liao:2016uis,Scholberg:2017czd}.   Roughly speaking, the magnitude of the Earth matter effect is proportional to the {\it difference} between the initial electron neutrino and non-electron neutrino spectra~\cite{Dighe:1999bi}. Since the difference of the spectra between electron and non-electron neutrinos turn out to be rather similar according to recent CCSN simulations the observation of the Earth matter effect seems to be more difficult than previously considered, unless the distance to CCSN from the Earth is significantly smaller than the typical distance of $ \sim$ 10 kpc~\cite{Borriello:2012zc}.  We will show that the decay effect can potentially enhance   the Earth matter effect even if the initial spectra of electron and non-electron neutrino species are very similar.  This is because the decay effect tends to enhance the difference by reducing the part of the spectra which is coming originally from the non-electron neutrino species, therefore, tends to enhance the Earth matter effect, as we will see later. 

The relevant quantity for the decay effect is $\tau E /(mD)$ where $\tau$, $m$ and $E$ are, respectively, proper lifetime, mass and energy of neutrino and $D$ is the distance between the source  (CCSN) and detection. For the distance $D=10$ kpc, and typical energy of CCSN neutrinos, $\sim 10$ MeV, we can roughly estimate the $\tau/m$ to have a large impact of decay just by considering the situation where $O(\tau E /(mD))\sim 1$ as
\begin{eqnarray}
  \frac{\tau}{m} \sim   \frac{D}{E} \sim 10^5
  \left[ \frac{D}{10\,\text{kpc}} \right]
  \left[ \frac{E}{10\,\text{MeV}} \right]^{-1} \ \frac{\text{s}}{\text{eV}}.
\end{eqnarray}

From the successful observations~\cite{Hirata:1987hu,Bionta:1987qt} of neutrinos, which were considered dominantly as $\bar{\nu}_e$, coming from SN1987A, located about 50 kpc away from the Earth, at least either $\tau_1/m_1$ or $\tau_2/m_2$ must be larger than $\sim 6 \times 10^5$ s/eV~\cite{Frieman:1987as}, where $\tau_i/m_i$ ($i=1,2,3$) imply the ratio of the lifetime to mass of $i$-th generation of neutrino. In this work we consider the cases where only one of $\tau_1/m_1$ or $\tau_2/m_2$ can be of order $\sim 10^5$ s/eV or smaller such that a significant fraction of $\bar{\nu}_e$ flux from SN1987A must have been arrived at the Earth in order to be consistent with the observed data~\cite{Frieman:1987as,Beacom:2002cb}.
Due to the uncertainty of the overall normalization of CCSN neutrino  fluxes, we assume that even a complete decay of either $\nu_1$ or  $\nu_2$  may not be excluded from the SN1987A neutrino data as well as from the future CCSN neutrino data and see if we can say something about the decay effect purely based on the Earth matter effect. For simplicity, we consider only the so called invisible decay of neutrinos  where the decay products are not observable.

The Earth matter effect manifest itself as the modulation in the observed CCSN neutrino spectra at the detectors~\cite{Dighe:2003jg}
which will be further influenced by the decay effect if exist. The modulation tends to be enhanced by the decay but not always as we will see later. In this work, we will discuss the possible impact of the neutrino decay on the Earth matter effect for JUNO-like, Hyper-Kamiokande-like and DUNE-like detectors, roughly mimicking, respectively, JUNO~\cite{An:2015jdp,JUNO:2021vlw},
Hyper-Kamiokande~\cite{Abe:2018uyc}  and DUNE~\cite{Abi:2020evt} detectors by taking into account only their main features, i.e., sizes and energy resolutions.


\section{Neutrinos from  core-collapse supernova}
\label{sec2}

Core collapse supernovae are the explosions that mark  the death of stars more massive than $\sim 8 M_{\odot}$, where $M_{\odot}$ is the mass of the Sun.  These explosions represent one of the most violent and energetic events in the Universe, where almost all the energy released by gravitational collapse is radiated away as neutrinos and anti-neutrinos of all flavors with energies of a few tens of MeV. The general feature of these neutrino emissions is rather well understood and can be divided roughly into three phases: neutronization burst,  accretion and cooling  of the proto-neutron star~\cite{Colgate:1966ax,Bethe:1990mw,Couch:2017}.  

\subsection{Un-oscillated neutrino  spectra}
\label{subsec2_1}

In a crude approximation, the newly formed proto-neutron star can be considered as a  black-body source for neutrinos of all flavors. Therefore, the energy distribution of these neutrinos can be well described  using a quasi-thermal spectrum. In this article, we will  adopt  a traditional  emission model, in which a CCSN  at a typical distance, $D = 10$ kpc,  emits a total gravitational binding energy of about $3 \times 10 ^{53}$ erg in approximately ten seconds. This energy, under  the assumption of energy equipartition, is shared among the 3 flavor neutrino and anti-neutrino species:  $\nu_{e}$, $\bar{\nu}_{e}$ and $\nu_x$ (i.e., $\nu_{\mu}$, $\nu_{\tau}$, and their antiparticles). 

In general, the luminosity $L_\nu$ and  the average  energy $\left< E_{\nu} \right>$  that characterize the energy spectrum of a given flavor   depend on the simulated CCSN explosion model  and  their evolution is also different at each CCSN phase.   However, the time-integrated energy  spectrum   obtained by these models can be well described by the following spectral function ~\cite{Keil:2002in} :
\begin{equation}
\label{eq:SN_spectra} 
F_{\nu}^0(E_\nu)    = \frac{L_\nu\,(\beta_\nu)^{\beta_\nu}}{4\pi\,D^2\,\left<E_\nu\right>^2\,\Gamma(\beta_\nu)}\,\left(\frac{E_\nu}{\left<E_\nu\right>} \right)^{\beta_\nu-1}\,\rm{exp}\left(-\frac{\beta_\nu\,E_\nu}{\left<E_\nu\right>} \right), 
\end{equation}
where  $\beta_\nu$ describes the deviation from a thermal distribution. Unless otherwise specified, as our default assumptions, we adopt CCSN neutrino parameters similar to the ones considered in several works, e.g. in~\cite{Borriello:2012zc,Machado:2012ee,Dighe:2003jg}: as representative average energies, we consider 13 MeV for the electron neutrino,  15 MeV  for the electron anti-neutrino and 18 MeV for  non-electron species, and $\beta_\nu = 4$ for all flavors. We call this set of parameters as the model A.  Later, in Sec.~\ref{sec:results} for some studies, we consider also 2 other different CCSN models for comparison.

\subsection{Adiabatic conversions in the mantle (MSW effect)}
\label{subsec2_2}
Neutrinos produced by CCSN interact  with the star constituents,  the result  is manifested as coherence or decoherence effects on their fluxes. When neutrinos experience coherent forward elastic scattering, the effect of the medium is described by an effective potential $V = \sqrt{2}\,G_F\,N_e$ that only the electron (anti-) neutrino can feel, where the common neutral current interactions for all flavors of neutrinos are ignored.  The potential is given in terms of the electron number density in the medium $(N_e)$ and the Fermi coupling constant $(G_F)$, a measure of the strength of the weak interaction at low energies. In the region where the matter density is very high, much higher than that in the center of Sun or Earth, the oscillation effect is strongly suppressed. However, if $V$ becomes comparable with the kinetic energy term $\Delta m^2/2E$, the oscillations can be significantly enhanced. This is the essence of the Mikheyev-Smirnov-Wolfenstein effect ~\cite{Wolfenstein:1977ue,Mikheev:1986gs}.

The evolution in time of the neutrino states at densities not very high (apart from the region very close to CCSN core) where neutrino-neutrino interactions can be ignored, is given in the ultrarelativistic limit through a Schr$\ddot{\text{o}}$dinger-like equation with  the effective Hamiltonian, in the flavor basis  written as  
\begin{eqnarray}
  H(x) =  \frac{1}{2E} 
   \left[ U \left(
  \begin{array}{ccc}
  0 & 0 & 0 \\
  0 & \Delta m_{21}^2 & 0 \\
  0 & 0 & \Delta m_{31}^2 \\
   \end{array}
   \right) U^\dag  +
   \left(
  \begin{array}{ccc}
  a(x) & 0 & 0 \\
   0 & 0 & 0\\
   0 & 0 & 0 \\
   \end{array}
   \right) 
   \right] ,
   \label{eqn:Hamiltonian} 
\end{eqnarray}
where $E$ is the neutrino energy, $\Delta m^2_{ij} \equiv m_i^2-m_j^2$ are the mass squared  differences of neutrinos, $m_i \,(i=1,2,3)$ being the neutrino masses, and $U$ is the standard mixing matrix which allows to describe the flavor eigenstates $\nu_{\alpha}\, (\alpha = e, \mu,  \tau)$ as mixtures of the  mass eigenstates $\nu_i \, (i = 1, 2, 3)$ as,
\begin{equation}
\ket{\nu_{\alpha}} = \sum_i U_{\alpha\,i}^{*} \ket{\nu_i}.
\label{eqn:neutrino_mix} 
\end{equation}

The term $a(x)$ in eq.~\eqref{eqn:Hamiltonian} takes account of the charged current interactions between electron (anti-) neutrinos and electrons:
\begin{eqnarray}
a(x) =  2E\,V(x) 
\approx  7.56 \times 10^{-8}\ \text{eV}^{2}
\left(
\frac{\rho(x)}{\text{g/cm}^3}
\right) 
\left(
\frac{E}{\text{MeV}}
\right) .
\label{eqn:potential} 
\end{eqnarray}

In this work we do not consider the so called ``collective'' oscillation effect caused by the neutrino-neutrino self-interactions~\cite{Pantaleone:1992eq} in a very dense region of CCSN core which may lead to very interesting phenomenon, see e.g.~\cite{Duan:2006jv,Duan:2006an,Hannestad:2006nj,Fogli:2007bk} for some earlier works and \cite{Duan:2010bg} for a review. The reason is for simplicity and because the effect is not yet very well understood and most importantly, the Earth matter effect, if detected, is expected to depend not so much on such effects.

\begin{figure}
\begin{center}
\vglue -0.5cm
\includegraphics[scale=0.65]{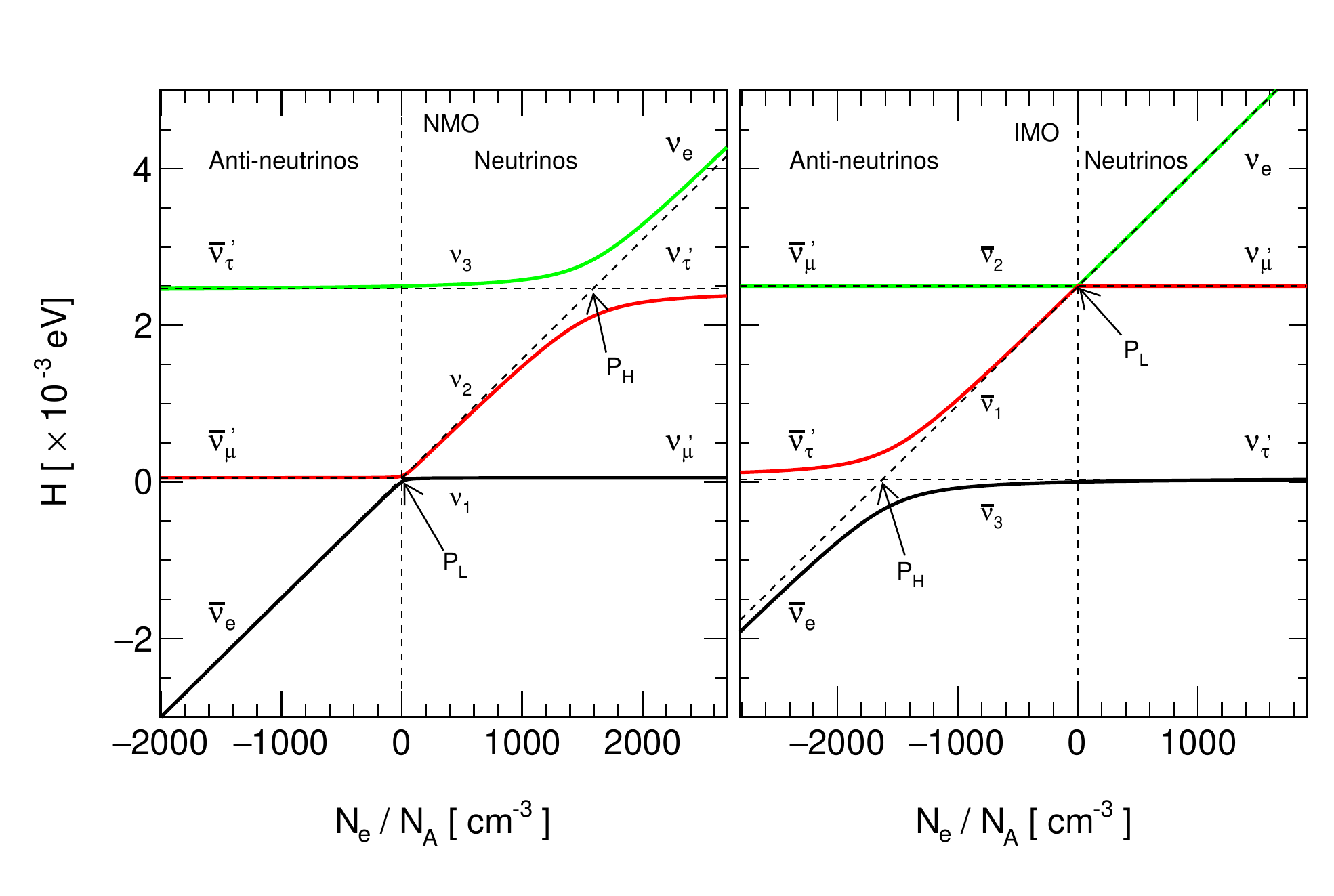}
\vglue -0.5cm
\caption{
Level crossing diagrams. Solid lines correspond to  eigenvalues of the effective Hamiltonian  and the dashed lines  to energies of flavor states. Anti-neutrinos are represented as neutrinos traveling through  the star medium with an effective negative potential, hence $N_e < 0$.
}
\label{figure:level-crossing-diagrams} 
\end{center}
\vglue -0.5cm
\end{figure}

By diagonalizing the Hamiltonian in the mass basis we obtain the mass eigenvalues in matter and therefore we can visualize its dependence on  $\rho(x)$, the  density in the medium at the position $x$.
In figure~\ref{figure:level-crossing-diagrams} we reproduce the eigenvalues of the Hamiltonian as a function of the electron number density or level crossing diagrams for the case where $m_1<m_2<m_3$, the normal mass ordering (NMO) shown in the left panel and the case where $m_3<m_1<m_2$ the inverted mass ordering (IMO) shown in the right panel, which can be found in several papers (see e.g.~\cite{Dighe:1999bi}). 
For the case of NMO, as we can see from the left panel of  figure~\ref{figure:level-crossing-diagrams},  electron neutrinos undergo two resonance points (H and L, corresponding to whether the density in that region is higher or lower) during the propagation through the mantle of the star whereas electron anti-neutrinos reach the surface without being influenced by any resonance point along their way.
On the other hand, for the case of IMO, electron neutrino and electron anti-neutrinos undergo, respectively, L and H resonance, as we can see in the right panel of figure~\ref{figure:level-crossing-diagrams}. 

We know from the LMA (Large Mixing Angle) solution to solar neutrino problem (see e.g., a review~\cite{Maltoni:2015kca}) that the L resonance is adiabatic, and also due to relatively large value of $\theta_{13}$ measured by reactors  \cite{DoubleChooz:2019qbj,Adey:2018zwh,Bak:2018ydk} we know that the passage through the H-resonance should be also  adiabatic, therefore, ignoring the possible collective oscillation effects in the dense region of CCSN as mentioned before, the CCSN neutrino spectra expected at the surface of the star are given by~ \cite{Dighe:1999bi}:      
\begin{subequations}
\label{eq:flux_NO}
\begin{align}
\label{eq:f_nue_NO}
F_{\nu_e} & = s^2_{13}F_{\nu_e}^0 + c^2_{13}F_{\nu_x}^0 \,, 
\\
\label{eq:f_nuebar_NO}
F_{\bar{\nu}_e}   & =  c^2_{12}c^2_{13}\,F_{\bar{\nu}_e}^0 
 + (1-c^2_{12}c^2_{13})\,F_{\nu_x}^0 \,,
\\
\label{eq:f_nux_NO}
2F_{\nu_x}  ( = F_{\nu_\mu} +F_{\nu_\tau} ) 
& = c^2_{13}F_{\nu_e}^0  + (1+s^2_{13})F_{\nu_x}^0  \,, 
\\
\label{eq:f_nuxbar_NO}
2F_{\bar{\nu}_x} ( = F_{\bar{\nu}_\mu}+F_{\bar{\nu}_\tau}) & 
= (1-c^2_{12}c^2_{13})F_{\bar{\nu}_e}^0 + (1+c^2_{12}c^2_{13})F_{\bar{\nu}_x}^0 \,, 
\end{align}
\end{subequations}
for NMO and, 
\begin{subequations}\label{eq:flux_IO}
\begin{align}
\label{eq:f_nue_IO}
F_{\nu_e} & = s^2_{12}c^2_{13}F_{\nu_e}^0 
+ (1-s^2_{12}c^2_{13})F_{\nu_x}^0 \,, 
\\
\label{eq:f_nuebar_IO}
F_{\bar{\nu}_e}  & =  s^2_{13}\,F_{\bar{\nu}_e}^0 + c^2_{13}\,F_{\bar{\nu}_x}^0\,,
\\
\label{eq:f_nux_IO}
2F_{\nu_x} ( = F_{\nu_\mu}+F_{\nu_\tau}) & =
 (1-s^2_{12}c^2_{13})F_{\nu_e}^0 
+ (1+s^2_{13}c^2_{13})F_{\nu_x}^0 \,, 
\\
\label{eq:f_nuxbar_IO}
2F_{\bar{\nu}_x} ( = F_{\bar{\nu}_\mu}+F_{\bar{\nu}_\tau}) & 
= c^2_{13}F_{\bar{\nu}_e}^0 + (1+s^2_{13})F_{\bar{\nu}_x}^0 \,, 
\end{align}
\end{subequations}
for IMO, where the notation $c_{ij} \equiv \cos \theta_{ij}$ and $s_{ij} \equiv \sin \theta_{ij}$ is used and the energy dependence in the spectra $F_{\nu}$ are implicit for simplicity. Here we treat the non-electron neutrino species, $\mu$ and $\tau$ neutrinos as a single species $\nu_x$ (or $\bar{\nu}_x$) as they have very similar interactions inside CCSN and their energy spectra are expected to be very similar.
Since the value of $s^2_{13}$ is rather small, $s^2_{13} \simeq 0.022$~\cite{Esteban:2020cvm,deSalas:2020pgw,Capozzi:2021fjo}, in this work, for simplicity and as a good approximation, we set $\theta_{13} =0$ in eqs.~\eqref{eq:flux_NO}  and  \eqref{eq:flux_IO}  and consider the following simplified CCSN neutrino spectra\footnote{Note that $\theta_{13}$ is not assumed to be zero for the neutrino evolution inside CCSN. Due to non-zero value of $\theta_{13}$, we assume that $H$ resonance is adiabatic. We are only ignoring the effect of $\theta_{13}$ after neutrinos exit from CCSN.}.
\begin{subequations}
\label{eq:flux_NO2}
\begin{align}
\label{eq:f_nue_NO2}
F_{\nu_e} & \approx F_{\nu_x}^0 \,, 
\\
\label{eq:f_nuebar_NO2}
F_{\bar{\nu}_e} & \approx  c^2_{12}\,F_{\bar{\nu}_e}^0 + s^2_{12}\,F_{\bar{\nu}_x}^0\,,
\\
\label{eq:f_nux_NO2}
2F_{\nu_x} & =
 F_{\nu_e}^0 + F_{\nu_x}^0 \,, 
\\
\label{eq:f_nuxbar_NO2}
2F_{\bar{\nu}_x}& 
\approx s^2_{12}F_{\bar{\nu}_e}^0 + (1+c^2_{12})F_{\bar{\nu}_x}^0 \,, 
\end{align}
\end{subequations}
for NMO, 
and  
\begin{subequations}
\label{eq:flux_IO2}
\begin{align}
\label{eq:f_nue_IO2}
F_{\nu_e} & \approx s^2_{12} F_{\nu_e}^0 + c^2_{12} F_{\nu_x}^0 \,, 
\\
\label{eq:f_nuebar_IO2}
F_{\bar{\nu}_e} & \approx \,F_{\bar{\nu}_x}^0\,,
\\
\label{eq:f_nux_IO2}
2F_{\nu_x} & =  c^2_{12}F_{\nu_e}^0 + (1+s^2_{12})F_{\nu_x}^0 \,, 
\\
\label{eq:f_nuxbar_IO2}
2F_{\bar{\nu}_x}& 
\approx F_{\bar{\nu}_e}^0 + F_{\bar{\nu}_x}^0 \,, 
\end{align}
\end{subequations}
for IMO.
On the way to the Earth the neutrino state lost coherence very quickly, so flavor oscillation ceases to occur and neutrinos arrive at the Earth as mass eigenstates. Therefore, the fluxes at the Earth are expected to be the same as that in eqs.~(\ref{eq:flux_NO2}) and (\ref{eq:flux_IO2}) .


\section{Invisible decay and Earth matter effects }
\label{sec3}
The successful observation of oscillation of neutrinos coming from the Sun, the atmosphere, reactors  and accelerators, confirms that at least two neutrinos are massive; therefore either Dirac or Majorana particles, where their mass term violates lepton number in the latter case. Among the logical possibilities, lepton number can be spontaneously broken globally and consequently a massless Nambu-Goldstone boson, associated with the neutrino mass generation, couples to neutrinos and makes possible the decay like $\nu_i \to \nu_j + \phi$ \cite{Chikashige:1980qk,Chikashige:1980ui,Schechter:1981cv,Gelmini:1983ea,Berezhiani:1987gf}. Where, the mass eigenstates ($\nu_i$) are usually referred to as parent neutrinos and the mass eigenstates ($\nu_j$) as daughter neutrinos. $\phi$ is the majoron, which is invisible in our detectors because it weakly couples with matter.  We note, however, that Dirac neutrinos also can decay, see e.g.~\cite{Acker:1991ej}.

According to the observability of decay products in the detector, we can classify neutrino decay into two types: (1) invisible decay, where neutrinos decay into some invisible states which could be light sterile neutrinos or active neutrinos which do not have enough energy to be detected and therefore not observables; and (2) visible decay, where the final states are detectable active neutrinos.

\subsection{Neutrino decay scenarios}
\label{subsec:nu-decay-scenarios}
In this work, we consider the case (1) where the daughter neutrinos are invisible. We do not consider any specific model of neutrino decay but just assume  that the decay product of neutrinos are not observable  from a phenomenological point of view. Furthermore, we consider the case where the decay rate of neutrino is  sufficiently small such that neutrinos decay only after they exit the star, on their way to Earth, and we ignore any decay effect of neutrinos inside the supernova.  In consequence, fluxes in eqs.~\eqref{eq:flux_NO} and \eqref{eq:flux_IO} can be modified substantially such that the observable Earth matter effect can be enhanced as we will see later. 

Let us start by analyzing the general case where the $\nu_i$ state can decay into some invisible (undetectable) states with the decay parameter $r_i$ ($i=1,2$ or 3), which are 
\begin{eqnarray}
r_i \equiv 1- \exp\left(-\frac{D}{E} \frac{m_i}{\tau_i} \right),
\label{eq:r-i}
\end{eqnarray}
where $m_i$ and $\tau_i$ are, respectively, the mass and the lifetime of the $i$-th neutrino mass eigenstate, $E$ is the neutrino energy
and $D$ is the distance traveled by neutrino.

In the presence of neutrino decay, the expected spectra at Earth are expressed as,  
\begin{subequations}
\label{eq:flux_NO_decay}
\begin{align}
\label{eqn3_1_1}
F_{\nu_e} & \approx (1-r_1c^2_{12} - r_2s^2_{12})F_{\nu_x}^0 \,, 
\\
\label{eqn3_1_2}
F_{\bar{\nu}_e} & \approx  
c^2_{12}(1-r_1)\,F_{\bar{\nu}_e}^0 + 
s^2_{12}(1-r_2)\,F_{\bar{\nu}_x}^0\,,
\\
\label{eqn3_1_3}
2F_{\nu_x} & =
(1-r_3) F_{\nu_e}^0 + (1-r_2)F_{\nu_x}^0 \,, 
\\
\label{eqn3_1_4}
2F_{\bar{\nu}_x}& 
\approx s^2_{12}(1-r_1)F_{\bar{\nu}_e}^0 + [1-r_3+c^2_{12}(1-r_2)]F_{\bar{\nu}_x}^0 \,, 
\end{align}
\end{subequations}
for NMO, and 
\begin{subequations}
\label{eq:flux_IO_decay}
\begin{align}
\label{eqn3_1_5}
F_{\nu_e} & \approx s^2_{12}(1-r_2) F_{\nu_e}^0 + c^2_{12}(1-r_1) F_{\nu_x}^0 \,, 
\\
\label{eqn3_1_6}
F_{\bar{\nu}_e} & \approx (1-c^2_{12}r_1-s^2_{12}r_2)
 \,F_{\bar{\nu}_x}^0\,,
\\
\label{eqn3_1_7}
2F_{\nu_x} & =  c^2_{12}(1-r_2)F_{\nu_e}^0 +
[1-r_3+s^2_{12}(1-r_1)]F_{\nu_x}^0 \,, 
\\
\label{eqn3_1_8}
2F_{\bar{\nu}_x}& 
\approx (1-r_1)F_{\bar{\nu}_e}^0 + (1-r_3)F_{\bar{\nu}_x}^0 \,, 
\end{align}
\end{subequations}
for IMO, where energy dependence is implicit.

Note that $r_i = 1$ means $\nu_i$ state disappear completely and the case $r_1=r_2=r_3=0$ recovers the standard CCSN neutrino
spectra without neutrino decay given in eqs.~(\ref{eq:flux_NO2}) and (\ref{eq:flux_IO2}). On the other hand, in the intermediate case, $r_i$ is expected to have energy dependence as shown in eq.~\eqref{eq:r-i}. Since we are mainly interested in the impact of decay for $\nu_e$ and $\bar{\nu}_e$ observations at the terrestrial detectors,  from now on we will ignore the decay of $\nu_3$ which have no impact for $\nu_e$ and $\bar{\nu}_e$ observations under the assumptions made in this work and, as typical representative cases, we consider the following 3 scenarios where $\nu_1$ is always stable but consider the possibility of decay (or not) of $\nu_2$: 
\begin{itemize}
\item [(S1)]
$\tau_{1,2}/m_{1,2}\gg 10^5$ s/eV
corresponding to the standard case (no decay) or  $r_1=r_2=0$. 
\item [(S2)] $\tau_{1}/m_{1}\gg 10^5$ s/eV with $\tau_{2}/m_{2}$ = $10^5$ s/eV, 
  or $r_1 = 0, r_2 \simeq 1-\exp(-[10\,\text{MeV}/E])$.
\item [(S3)]
 $\tau_{1}/m_{1} \gg 10^5$ s/eV but $\tau_{2}/m_{2}\ll 10^5$ s/eV, 
  or $r_1 = 0, r_2 \simeq 1$.
\end{itemize}

One might consider that the case like S3 would lead to a significant reduction of  CCSN $\nu_e$ and $\bar{\nu}_e$ at Earth and can be easily identified the decay effect but the total number of neutrino events depends on CCSN model parameters (such as average neutrino energies, luminosities, the spectrum parameter $\beta$) which are not very well known and the distance to CCSN which may not be well determined. Therefore, in this work we assume that just based on the detected number of total  neutrino events, it would be difficult
to identify the decay effect.  

\subsection{Earth matter effects in presence of neutrino decay}
CCSN neutrino wave packets lose coherence quickly on their way to the Earth, hence neutrinos arrive at the surface as incoherent
mixtures of fluxes of the mass eigenstates ($\nu_1,\, \nu_2$ and $\nu_3$). These eigenstates do not coincide with the eigenstates of the Hamiltonian in matter. Therefore,  inside the Earth  coherence can be partially  restored and  as a consequence, flavor oscillations are obtained~\cite{Kersten:2015kio}.  Here after,  we consider only the electron type neutrinos, $\nu_e$ and $\bar{\nu}_e$,  as they are the main neutrinos which affect the main observables of detectors like JUNO, Hyper-K and DUNE.

We expect that the CCSN neutrino spectra given in eqs.~(\ref{eq:flux_NO_decay}) and (\ref{eq:flux_IO_decay})  will get modified as the  neutrinos propagate through the Earth according to
\begin{subequations}\label{flux_NO}
\begin{align}
\label{eqn3_2_1}
F_{\nu_e}^\oplus & \approx (1-r_1\, P_{1e}^\oplus - r_2\, P_{2e}^\oplus)F_{\nu_x}^0 \,, 
\\
\label{eqn3_2_2}
F_{\bar{\nu}_e}^\oplus & \approx  
\bar{P}_{1e}^\oplus(1-r_1)\,F_{\bar{\nu}_e}^0 + 
\bar{P}_{2e}^\oplus(1-r_2)\,F_{\bar{\nu}_x}^0\,,
\end{align}
\end{subequations}
for NMO, and  
\begin{subequations}\label{flux_IO}
\begin{align}
\label{eqn3_2_3}
F_{\nu_e}^\oplus & 
\approx P_{2e}^\oplus(1-r_2) F_{\nu_e}^0 + P_{1e}^\oplus(1-r_1) F_{\nu_x}^0 \,, 
\\
\label{eqn3_2_4}
F_{\bar{\nu}_e}^\oplus & \approx 
(1-\bar{P}_{1e}^\oplus\, r_1-\bar{P}_{2e}^\oplus\, r_2)
 \,F_{\bar{\nu}_x}^0\,,
\end{align}
\end{subequations}
for IMO, where $P_{ie}^\oplus \equiv P({\nu}_i \to {\nu}_e)^\oplus$ ($\bar{P}_{ie}^\oplus \equiv P({\bar{\nu}}_i \to {\bar{\nu}}_e)^\oplus$) is the probability that $\nu_i$ ($\bar{\nu}_i$) state will be detected as $\nu_e$ ($\bar{\nu}_e$) state after traversing the Earth, which depend on neutrino energy, the distance ($L$) traveled by neutrino inside the Earth and mixing parameters, $\Delta m^2_{21}$ and $ \theta_{12}$.  We  compute these  probabilities by numerically solving the evolution equation of neutrinos using the PREM model~\cite{PREM} for the matter density profile for the Earth. For an analytic treatment applicable to the oscillation of CCSN neutrinos passing throught the Earth matter, see e.g. ref.~\cite{Ioannisian:2004vv}.

The impact of the Earth matter through the regeneration effect in the oscillations can be characterized by $\Delta F_{\nu_e}$ or $\Delta F_{\bar{\nu}_e}$, the difference between the CCSN neutrino spectra with (and without) Earth matter effect $F_\nu^\oplus$ ($F_\nu$) as, 
\begin{subequations}
\label{eq:Delta_F_Earth_NO}
\begin{align}
\label{eqn3_2_5}
\Delta F_{\nu_e} \equiv F_{\nu_e}^\oplus - F_{\nu_e} &= f_\text{reg}(r_1-r_2)F_{\nu_x}^0 ,
\\
\label{eqn3_2_6}
\Delta F_{\bar{\nu}_e} \equiv 
F_{\bar{\nu}_e}^\oplus - F_{\bar{\nu}_e} & = 
\bar{f}_\text{reg}[(1-r_2)F_{\bar{\nu}_x}^0-(1-r_1)F_{\bar{\nu}_e}^0 ],
\end{align}
\end{subequations}
for NMO and 
\begin{subequations}
\label{eq:Delta_F_Earth_IO}
\begin{align}
\label{eqn3_2_7}
\Delta F_{\nu_e} \equiv F_{\nu_e}^\oplus - F_{\nu_e} & = 
f_\text{reg}[(1-r_2)F_{{\nu}_e}^0-(1-r_1)F_{{\nu}_x}^0 ],
\\
\label{eqn3_2_8}
\Delta F_{\bar{\nu}_e} \equiv 
F_{\bar{\nu}_e}^\oplus - F_{\bar{\nu}_e} & = 
\bar{f}_\text{reg}(r_1-r_2)F_{\bar{\nu}_x}^0,
\end{align}
\end{subequations}
for IMO. In the above equations, $f_\text{reg}$ and $\bar{f}_\text{reg}$ are, respectively, Earth regeneration factors \cite{Lunardini:2001pb} for $\nu$ and $\bar{\nu}$ given by 
\begin{subequations}
\begin{align}
\label{eqn2_2_1}
f_\text{reg} \equiv P_{2e}^\oplus - s^2_{12} = c^2_{12} - P_{1e}^\oplus, \\
\label{eqn2_2_2}
\bar{f}_\text{reg} \equiv  \bar{P}_{2e}^\oplus - s^2_{12} 
= c^2_{12} - \bar{P}_{1e}^\oplus.
\end{align}
\end{subequations}

For the case where the Earth matter density can be considered as constant,  these regeneration factors are approximately given by~\cite{Dighe:2003jg}
\begin{subequations}
\begin{align}
f_\text{reg} \approx \sin 2 \theta^\oplus_{e2} \sin(2 \theta^\oplus_{e2}-2 \theta_{12})
\sin^2\left(12.5 \frac{\Delta m^2_\oplus L}{E}\right),
\\
\bar{f}_\text{reg} \approx \sin 2 \bar{\theta}^\oplus_{e2} \sin(2 \bar{\theta}^\oplus_{e2}-2 \theta_{12})
\sin^2\left(12.5 \frac{\overline{\Delta m^2_\oplus} L}{E}\right),
\end{align}
\label{eq:f_reg}
\end{subequations}
where $\theta^\oplus_{e2}$ ($\bar{\theta}^\oplus_{e2}$) is the mixing angle between $\nu_e$ and $\nu_2$  ($\bar{\nu}_e$ and $\bar{\nu}_2$) in the Earth matter whereas $\Delta m^2_\oplus$ ($\overline{\Delta m^2_\oplus}$) is the mass squared difference between
$\nu_1$ and $\nu_2$ ($\bar{\nu}_1$ and $\bar{\nu}_2$) given in units of $10^{-5}$ eV$^2$ and $L$ is given in units of 1000 km. 

Note, however, that $\Delta F_{\nu_e}$ and $\Delta F_{\bar{\nu}_e}$ are not directly observable because for a given detector location, CCSN neutrinos either pass or not the Earth so that in practice we can not take the difference of the 2 neutrino fluxes with and without Earth matter effect but $\Delta F$'s are useful quantities to have some idea.

For the scenario without any decay effect (S1), we recover the standard  Earth matter effects for CCSN neutrinos which
exist only for $\nu_e$ for IMO and $\bar{\nu}_e$ for NMO~\cite{Dighe:1999bi}, but with the effects of neutrino decay, the Earth matter effects can manifest for both $\nu_e$ and $\bar{\nu}_e$ at the same time for both mass orderings; roughly speaking, imitating  the results that will be obtained if the high resonance (in the star) were partially adiabatic or completely non-adiabatic, as reported in table 2 of~\cite{Dighe:1999bi}. This implies that if the Earth matter effects will be observed in both $\nu_e$ and $\bar{\nu}_e$ events, this can be an indication of neutrino decay since the high resonance is expected to be very adiabatic in the standard oscillation scenario inside CCSN~\cite{Dighe:1999bi}. 

For the special case where  $\nu_2$ decays completely, $(r_1,r_2) = (0,1)$, we have some simplified relations as, 
\begin{subequations}
\label{eq:Delta_F_Earth_NO_IO}
\begin{align}
\label{eqn2_2_1}
\Delta {F_{\nu_e}}^\text{\tiny NMO}_{(0,1)} 
=  \Delta {F_{\nu_e}}^\text{\tiny IMO}_{(0,1)} 
 &=  -f_\text{reg}F_{\nu_x}^0 ,
\\
\label{eqn2_2_2}
\Delta {F_{\bar{\nu}_e}}^\text{\tiny IMO}_{(0,1)} 
& = 
-\bar{f}_\text{reg}F_{\bar{\nu}_x}^0 ,
\\
\Delta {F_{\bar{\nu}_e}}^\text{\tiny NMO}_{(0,1)} 
& = 
-\bar{f}_\text{reg}F_{\bar{\nu}_e}^0,
\end{align}
\end{subequations}
where superscripts NMO (IMO) implies normal (inverted) mass ordering. Note that in this case even if all the neutrino species have the same spectra, or even if  $F_{{\nu}_e}^0$  = $F_{\bar{\nu}_e}^0$ =   $F_{\bar{\nu}_x}^0$  we expect that  $\Delta {F_{\nu_e}} \ne 0$ and $\Delta {F_{\bar{\nu}_e}} \ne 0$ inducing some Earth matter effect which would not happen for the standard neutrinos without decay.
\begin{figure}
\begin{center}
\includegraphics[scale=0.62]{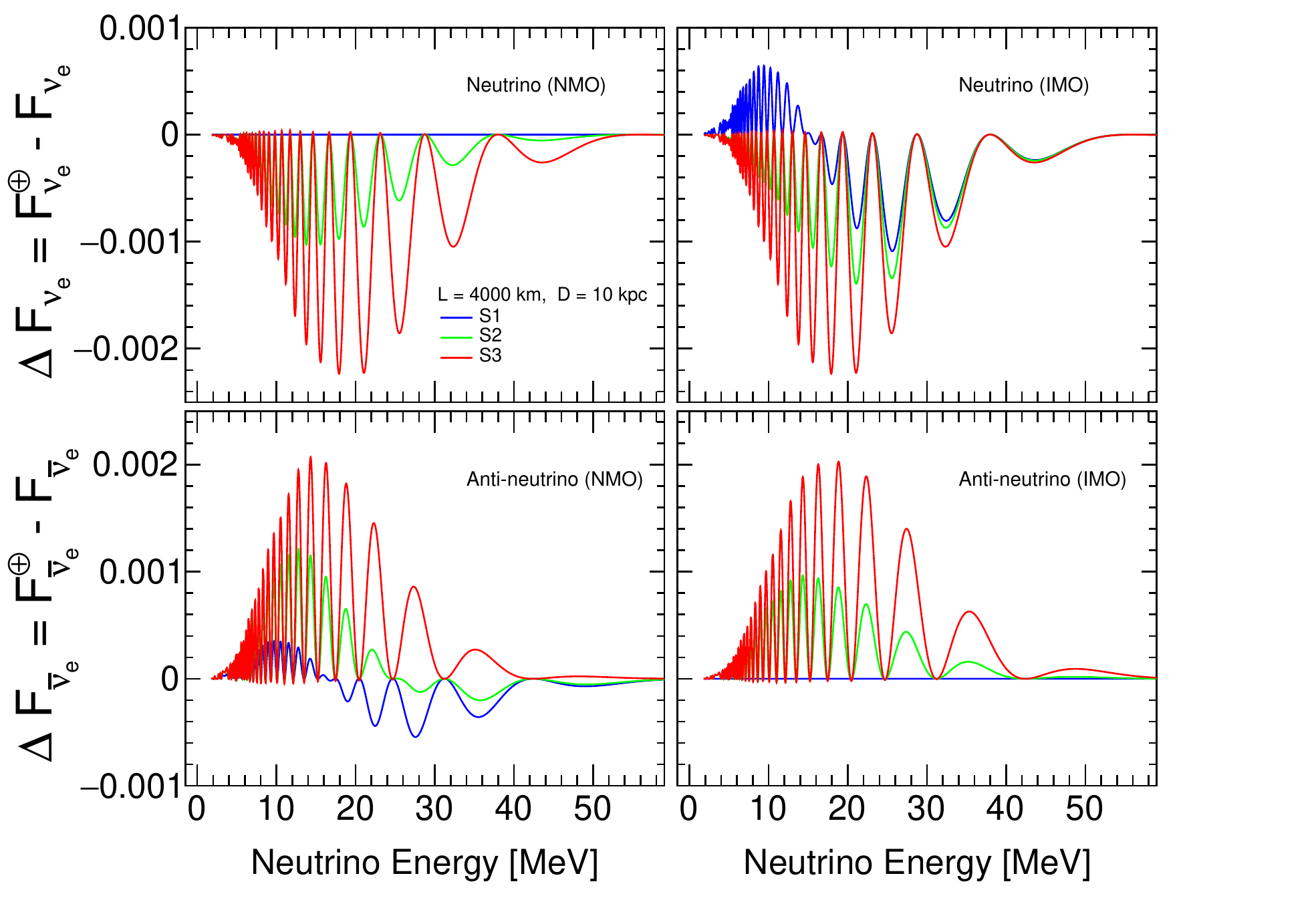}
\vglue -0.1cm
\caption{
  $\Delta F$ defined in eqs.~(\ref{eq:Delta_F_Earth_NO}) and~(\ref{eq:Delta_F_Earth_IO}) are shown as a function of neutrino energy computed for $L=4000$ km (the distance traveled by CCSN neutrinos inside the Earth) for $F_{\bar{\nu}_e}^\oplus$ or $F_{{\nu}_e}^\oplus$, for neutrino (upper panels) and anti-neutrinos (lower panels) for normal mass ordering (left panels) and inverted mass ordering (right panels).   We have considered the scenarios S1, S2 and S3 indicated, respectively, by the solid blue, green and red curves.
  To eliminate the dependence of the distance to supernova, we have normalized the spectra such that $\int F_\nu^0(E_\nu)\, dE_\nu = 1$ for any species of neutrino. }
\label{figure:Delta_F_4000km}
\end{center}
\vglue -0.5cm
\end{figure}

In figure~\ref{figure:Delta_F_4000km} we show the difference between the CCSN neutrino spectra with and without Earth matter effect as a function of neutrino energy computed by assuming CCSN parameters given in  section~\ref{sec2} for $L=4000$ km,  for neutrino (upper panels) and anti-neutrinos (lower panels) for NMO  (left panels) and IMO (right panels).

We have considered the scenarios S1, S2 and S3 indicated, respectively, by the solid blue, green and red curves.  We first observe that in the case of no neutrino decay (S1) there is no Earth matter effects for $\nu_e$ $(\bar{\nu}_e)$ when the mass ordering is normal (inverted) in consistent with the expectation discussed in \cite{Dighe:1999bi}.
However, in the presence of neutrino decay, Earth matter effect can be present simultaneously for $\nu_e$ and $\bar{\nu}_e$ for both mass orderings.  We can see that by comparing the green curve (the case S2) and the red curve (the case S3, 100\% decay) that the Earth matter effects get larger for larger decay rate, and when the decay rate is large (when more than $\sim$ 60\,--\,70\% of $\nu_2$ decay) the impact is similar for NMO and IMO.

\begin{figure}[H]
\begin{center}
\vglue -0.2cm
\includegraphics[scale=0.62]{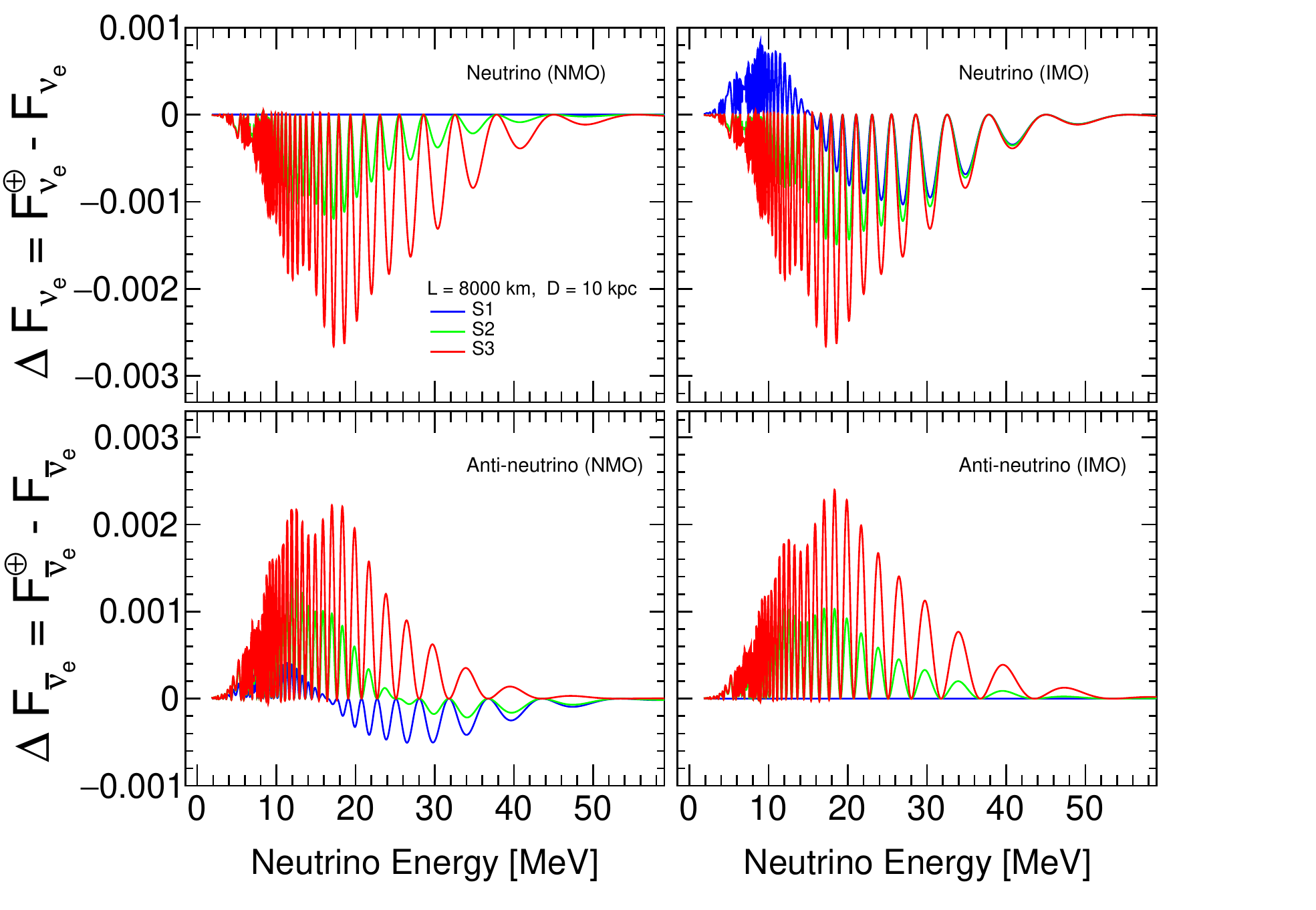}
\vglue -0.5cm
\caption{
  $\Delta F$ defined in eqs.~(\ref{eq:Delta_F_Earth_NO})  and~(\ref{eq:Delta_F_Earth_IO}) are shown as a function of neutrino energy computed  for $L=8000$ km (the distance traveled by CCSN neutrinos inside the Earth) for $F_{\bar{\nu}_e}^\oplus$ or $F_{{\nu}_e}^\oplus$,  for neutrino (upper panels) and anti-neutrinos (lower panels) for NMO (left panels) and IMO (right panels). We have considered the scenarios S1, S2 and S3 indicated, respectively, by the solid blue, green and red curves.  To eliminate the dependence of the distance to supernova, we have normalized the spectra such that $\int F_\nu^0(E_\nu)\, dE_\nu = 1$ for any species of neutrino. }
\label{figure:Delta_F_8000km}
\end{center}
\vglue -0.5cm
\end{figure}

\begin{figure}[H]
\begin{center}
\vglue -0.2cm
\includegraphics[scale=0.62]{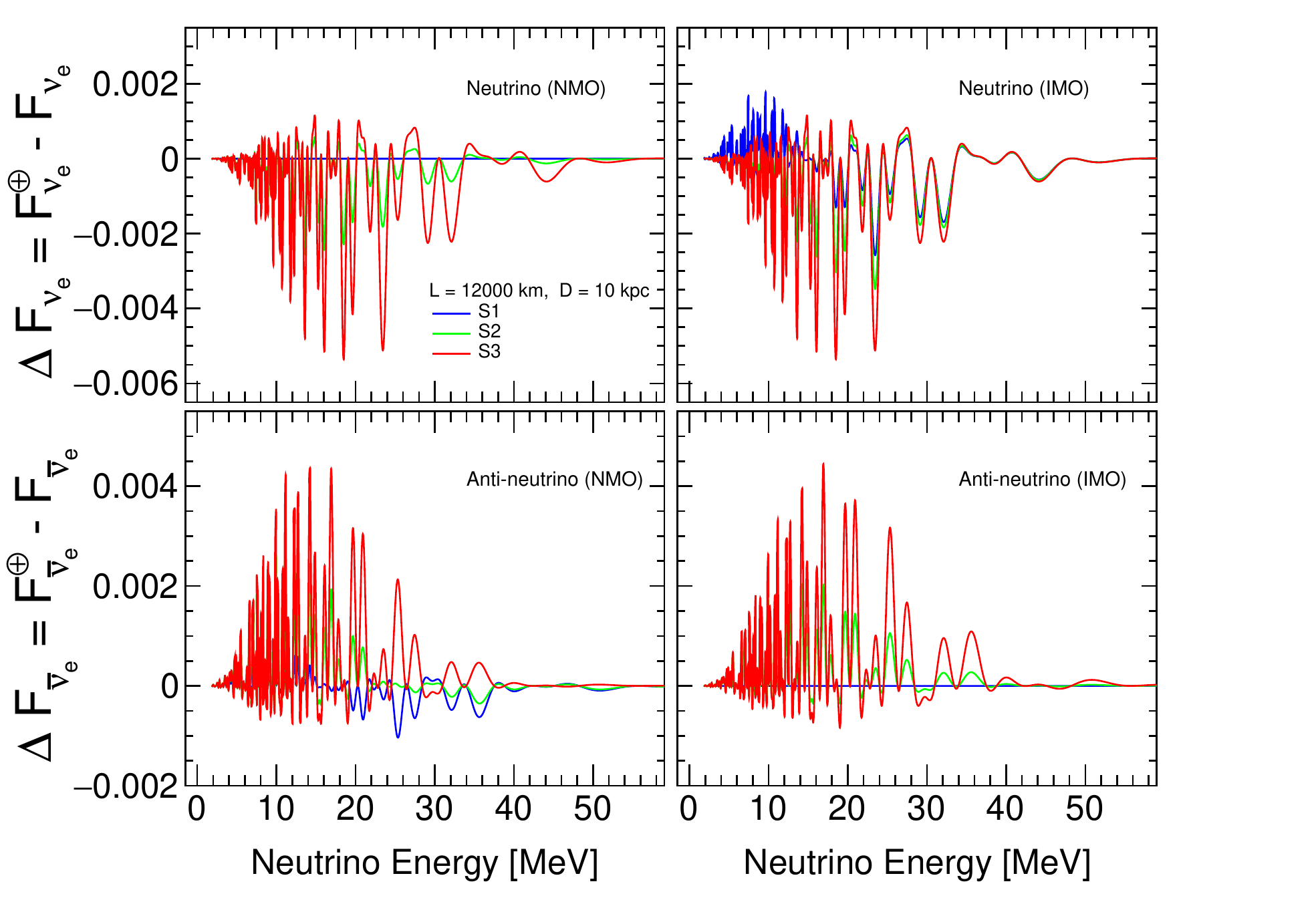}
\vglue -0.5cm
\caption{
  $\Delta F$ defined in eqs.~(\ref{eq:Delta_F_Earth_NO})  and~(\ref{eq:Delta_F_Earth_IO}) are shown as a function of neutrino energy computed  for $L=12000$ km (the distance traveled by CCSN neutrinos inside the Earth) for $F_{\bar{\nu}_e}^\oplus$ or $F_{{\nu}_e}^\oplus$, for neutrino (upper panels) and anti-neutrinos (lower panels) for NMO (left panels) and IMO (right panels). We have considered the scenarios S1, S2 and S3 indicated, respectively, by the solid blue, green and red curves.  To eliminate the dependence of the distance to supernova, we have normalized the spectra such that $\int F_\nu^0(E_\nu)\, dE_\nu = 1$ for any species of neutrino. }
\label{figure:Delta_F_12000km}
\end{center}
\vglue -0.5cm
\end{figure}

In figure~\ref{figure:Delta_F_8000km} we show the similar results shown in figure~\ref{figure:Delta_F_4000km} but for the case where $L=8000$ km and in figure~\ref{figure:Delta_F_12000km} for the case where $L=12000$ km.  The results for $L=8000$ km (shown in figure~\ref{figure:Delta_F_8000km}) are qualitatively similar  to that for $L=4000$ km (shown in figure~\ref{figure:Delta_F_4000km}) because for both cases neutrinos pass only the Earth mantle.   

The main difference between the cases of $L=4000$ km and 8000 km is that for larger value of $L$, the the modulation frequency become larger, which is expected  by eq.~\eqref{eq:f_reg}. On the other hand, the case for $L=12000$ km show different features, irregular spike-like structures, which is caused by the fact that neutrinos
pass the Earth core.


\section{Enhancing sensitivity  to Earth matter effects detection by a single detector with $\nu_2\, (\bar{\nu}_2)$ decay}
\label{sec:results}
A way of identifying the presence of matter effects in fluxes of neutrinos that pass through the Earth on their way to the detector is to observe the modulations  that appear in their energy spectrum which is caused by the regeneration effects. However, recent studies indicate that these features are expected to be difficult to be observed because of the similarity of average energies between electron and other flavors of neutrinos unless the CCSN is rather close to Earth, say less than a few kpc away~\cite{Borriello:2012zc}.

SN1987A in the Large Magellanic Cloud, one of the satellite galaxies of the Milky Way,  is the unique evidence  we have of the CCSN emission spectra but because of the small detected number of events, there is not enough precision to determine what mixture of the initial $\bar{\nu}_e$  and $\nu_x$ led to the observed spectrum. This statistical limitation in the determination of the $\bar{\nu}_e$ spectrum  currently allows us  to contemplate new scenarios that in a sub-dominating  form allow a clearer determination of the Earth matter effects. Compared to that epoch, we have currently larger number and better capability of neutrino detectors in operation and we have more chance to confirm or discard these new scenarios if neutrinos from a nearby CCSN are detected by these detectors in the near future. 

Here, we  are looking towards the possibility if invisible neutrino decay can lead to situations where the effective difference between these primary  fluxes is large enough to make the Earth matter effect observable at a single detector like JUNO~\cite{An:2015jdp}, Hyper-Kamiokande~\cite{Abe:2018uyc} or DUNE~\cite{Abi:2020evt}. We stress that as mentioned in the introduction, in this paper,
we consider only the main features of these detectors,  namely, sizes and energy resolutions, to approximately mimic them, which  will be sufficient for our purpose.  Hereafter whenever we discuss our results for these detectors, we mean approximated JUNO-like, Hyper-Kamokande-like and DUNE-like detectors, and for simplicity we will be mostly omitting to write ``-like''  explicitly. 


\subsection{The decay effect impact for JUNO}

First let us discuss the impact for the JUNO(-like) detector.  This 20 kt liquid scintillator detector, under construction in China, has as its main aim to determine neutrino mass ordering by measuring very precisely the energy spectrum of reactor electron anti-neutrinos (mainly due to its good energy resolution), which also allows JUNO to perform precise determination of other mixing parameters as well. Among other capabilities,  JUNO can detect CCSN neutrinos~\cite{An:2015jdp,Lu:2016ipr,Lu:2014zma} mainly via the inverse beta decay (IBD) reaction and also by some other channels such as $\nu-e^-$ elastic scattering and charged current reactions on  $^{12}$C, giving much less contributions.
For simplicity we consider only IBD reaction and ignore others\footnote{The elastic neutrino-proton scattering gives relatively large $\sim O(10^3)$ number of events for the 10 kpc distance to CCSN~\cite{An:2015jdp,JUNO:2021vlw} but this channel would not be sensitive to the Earth matter effect (or to the oscillation effect among active neutrino flavors) as it is induced by neutral current reactions common for all flavor.}. 

At the IBD channel, a CCSN electron anti-neutrino interact with  a free proton in the liquid scintillator, creating  a positron and a neutron. The positron quickly  annihilates with an electron and deposits at the detector a prompt signal, $E_{v} = E_{e^+} + m_e$ called visible energy. The neutron scatters  through the detector and is later captured by a proton ($\sim 200$ $\mu$s after its creation). The coincidence of the prompt and  delayed signal significantly reduces backgrounds~\cite{An:2015jdp,JUNO:2021vlw}. 

JUNO is expected to register roughly about 5000-6000 IBD CCSN neutrino events~\cite{An:2015jdp,JUNO:2021vlw} without decay effect, though the precise number of events depend on CCSN model parameters and the distance to CCSN, that  we can distribute in bins with the purpose of obtaining relevant information for our work. The number of events in the $i$-th bin can be computed as 
\begin{equation}
(\mathcal{N}_{\bar{\nu}_e})_i =  N_T \int_{E_i^{v} - \Delta E_i^{v}/2}^{E_i^{v} + \Delta E_i^{v}/2}\,dE_{v} \int_{E_{\bar{\nu}_e}^{th}}^\infty dE_{\bar{\nu}_e}\,  F_{\bar{\nu}_e}^\oplus (E_{\bar{\nu}_e})\,\sigma_{\bar{\nu}_e\,p}(E_{\bar{\nu}_e})\,\mathfrak{R}(E_{v}; E_{\bar{\nu}_e}, \delta E_{v})\,.
\label{eqn4_1}
\end{equation}

Here,  $F_{\bar{\nu}_e}^\oplus$  is given by eqs.~(\ref{flux_NO}) and (\ref{flux_IO}). It includes all the flavor conversions that have taken place from its creation to its detection, and also the flux transformation due to $\bar{\nu}_2$  decay.

The IBD cross section $\sigma_{\bar{\nu}_e\,p}(E_{\bar{\nu}_e})$ is implemented using the formula  found in ref.~\cite{Strumia:2003zx}, $N_T \sim 1.46\times10^{33}$ is the number of free protons at the detector, $E_{\bar{\nu}_e}^{\text{th}} = 1.806$ MeV is the energy threshold of the reaction and $\mathfrak{R}(E_{v}; E_{\bar{\nu}_e}, \delta E_{v})$ is the normalized Gaussian smearing function which takes into account  the  energy resolution of the detector. We define this function to be:
\begin{equation}
\mathfrak{R}(E_{v}; E_{\bar{\nu}_e}, \delta E_{v}) = \frac{1}{\sqrt{2\pi}\delta E_{v}}\, \exp{\left[-\frac{1}{2}\left(\frac{E_{v} - E_{\bar{\nu}_e} + 0.782 \,\text{MeV}}{\delta E_{v}}\right)^2\right]},
\label{eq:energy-resolution-function}
\end{equation}
where  $\delta E_{v}/\text{MeV}  =   3\%\,\sqrt{E_\nu/\text{MeV}}$ is the energy resolution of the detector.  For simplicity,  in our JUNO-like setting, we have ignored the energy nonlinearity of the liquid scintillator detector since it is expected to be small for event with larger than 2 MeV energy, which justifies our simplified treatment for our purpose.
As long as the recoil energies of nucleons are negligible, the neutrino energy $(E_{\bar{\nu}_e})$ and the positron energy  $(E_{e^+})$  can be related as  $E_{e^+} = E_{\bar{\nu}_e} - (m_n - m_p) \approx E_{\bar{\nu}_e} - 1.293$ MeV and therefore, the visible energy can be approximated to  $E_{v} \approx E_{\bar{\nu}_e} - 0.782 $ MeV.

\begin{figure}
\begin{center}
\vglue -0.2cm
\includegraphics[scale=0.65]{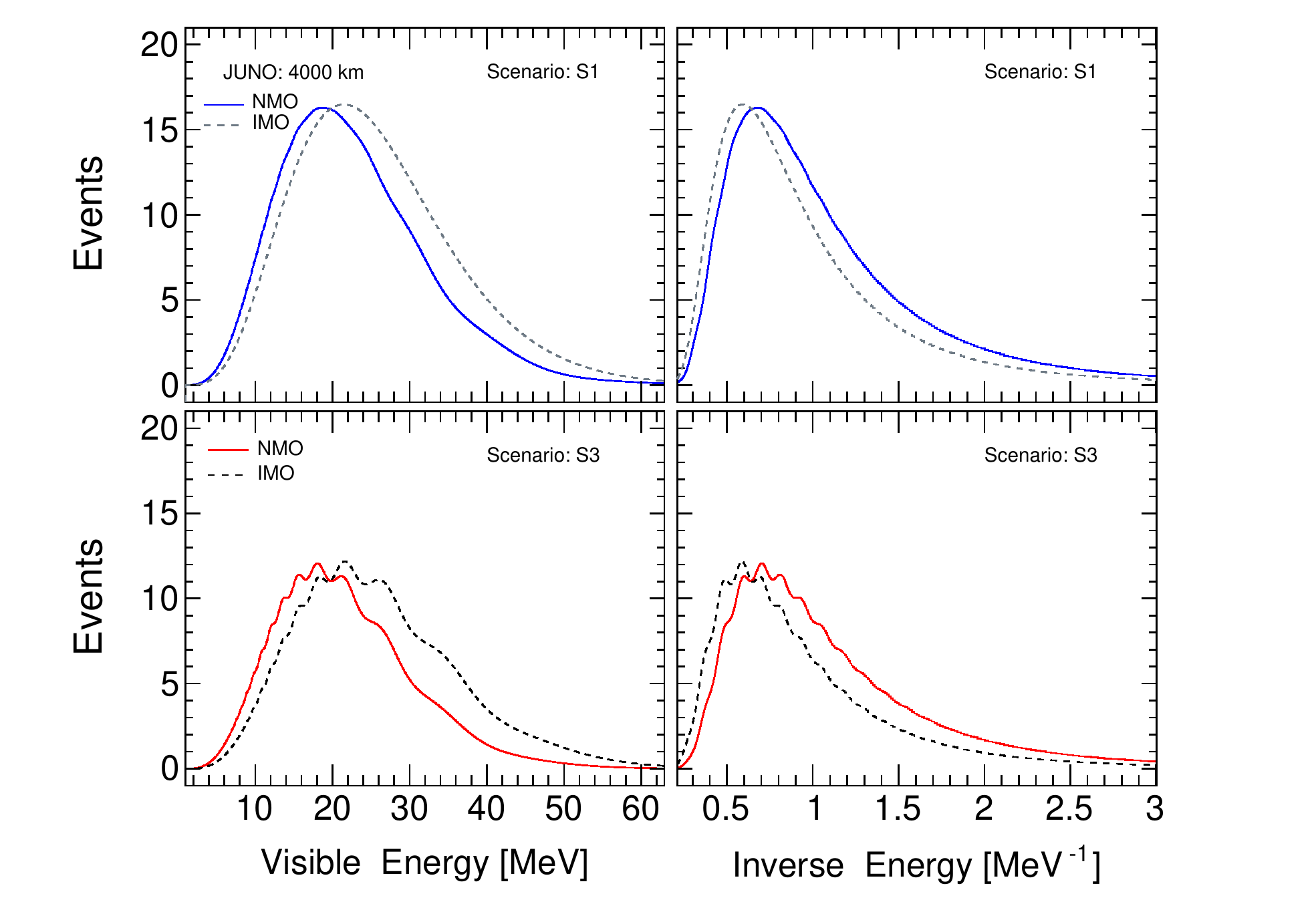}
\vglue -0.4cm
\caption{
  Visible (left panels) and inverse (right panels) energy distribution of IBD events expected at our JUNO-like detector for $L = 4000$ km. The solid blue and red histograms refer specifically to S1 (no decay) and S3 (100\% decay of $\nu_2$)  scenarios computed for NMO.
  The dashed gray and black histograms refer to the same decay scenarios but for the case of IMO.  }
\label{figure:event_distr_juno_gg_10kpc_4000km}
\end{center}
\vglue -0.5cm
\end{figure}

In figure~\ref{figure:event_distr_juno_gg_10kpc_4000km}, we present, as an illustrative example, the visible (left panels) and the inverse (right panels) energy distribution of the number of IBD events expected at our JUNO-like detector for $L = 4000$ km. The solid blue and red histograms refer specifically to S1 (no decay) and S3 (100\% decay of $\nu_2$) scenarios computed for NMO. We note that as discussed in \cite{Dighe:2003jg} the modulations on the inverse energy spectrum caused by the Earth matter effect  are roughly equally spaced which is expected by eq.~\eqref{eq:f_reg}. The dashed gray and black histograms refer to the same decay scenarios but for the case of IMO.  

We can highlight two important characteristics in the figure. The first comes from the comparison of the event distribution  for the different mass orderings; as a result we note that IMO spectra are harder than the NMO ones due to the MSW effect inside CCSN which leads to larger observable matter effects in IMO due to higher statistics.

The second is due to neutrino decay, whose presence  considerably alters the form of these histograms. Particularly, when $\bar{\nu}_2$ decay completely we can clearly appreciate the appearance of modulations. Despite the reduction of the total number of events due to decay, stronger modulations can potentially allow us  to detect more easily the Earth matter effects compared to the case without decay.
As considered in the previous works~\cite{Dighe:2003jg,Borriello:2012zc,Liao:2016uis}  we can identify the frequency that characterize these modulations by taking the Fourier transform of the inverse energy spectrum showed in figure~\ref{figure:event_distr_juno_gg_10kpc_4000km} (See appendix for details of how calculating  this transform). The observability of one or three peaks in the power spectrum, one for each important jump\footnote{When neutrinos only traverse the mantle, they encounter only one jump in the density: the difference between the vacuum and mantle densities (by regarding that the Earth atmosphere is vacuum). But, when neutrinos also go through the core they encounter two additional jumps corresponding to the differences between the mean densities in the mantle-core and core-mantle boundaries.} in the density that neutrinos encounter
when they travel inside the Earth,  would be a clear sign of the presence of matter effects.                              

\begin{figure}[H]
\begin{center}
\includegraphics[scale=0.64]{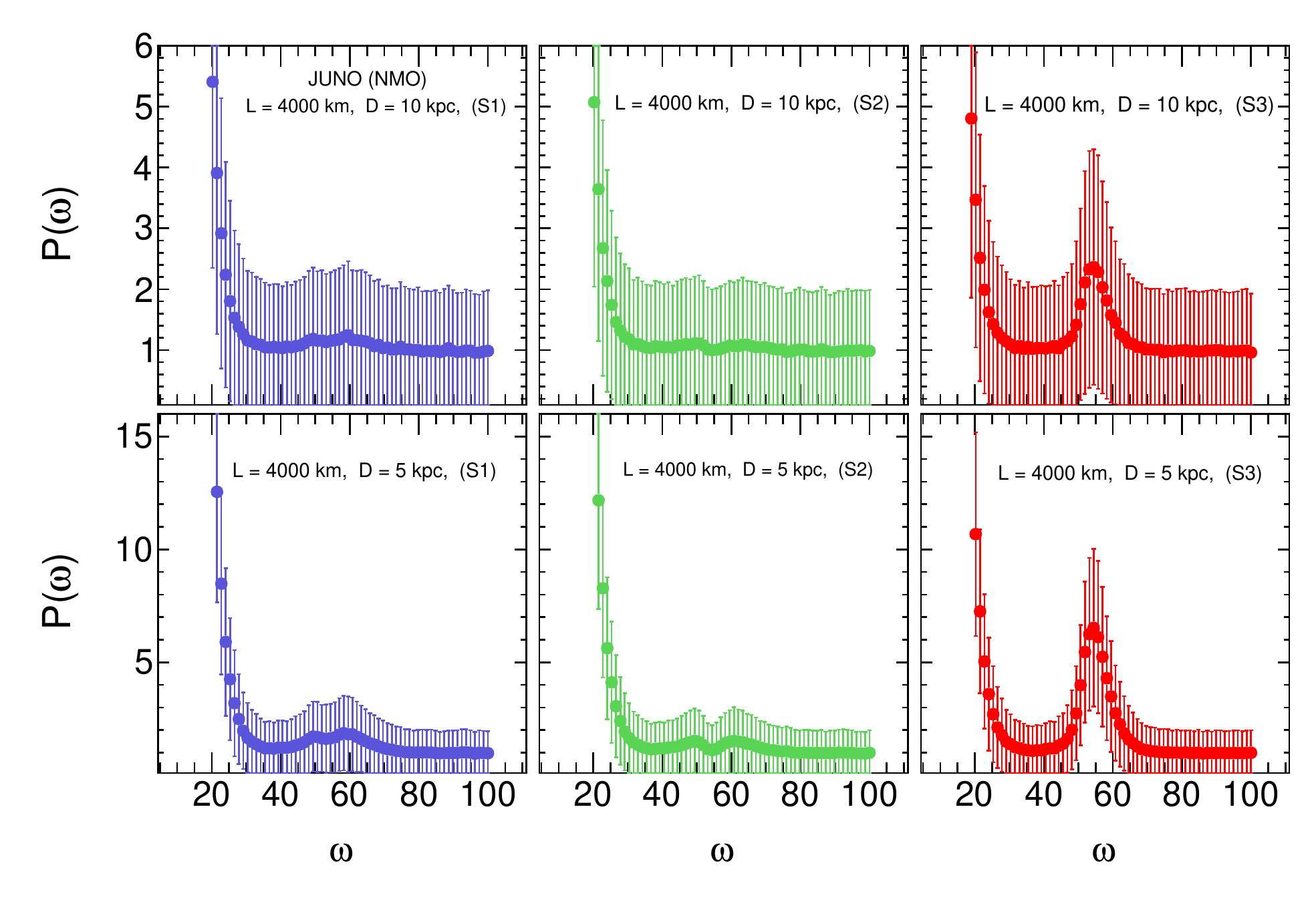}
\vglue -0.5cm
\caption{
  Averaged power spectrum over 1000 MC (Monte Carlo) run samples for $L = 4000$ km  and normal mass ordering.  We present these results for a CCSN distance of 10 kpc (upper panels) and 5 kpc (lower panels)  for the 3 scenarios, S1 (no decay), S2 ($\sim$ 50\% decay) and S3 (100\% decay) described in the subsection~\ref{subsec:nu-decay-scenarios}.  As error in the measurement, the standard deviation given in the fit has been considered and indicated by vertical error bars.  }
\label{figure:pow_spectrum_juno_gg_4000km_NO}
\end{center}
\vglue -0.5cm
\end{figure}

\begin{figure}[H]
\begin{center}
\vglue -0.5cm
\includegraphics[scale=0.64]{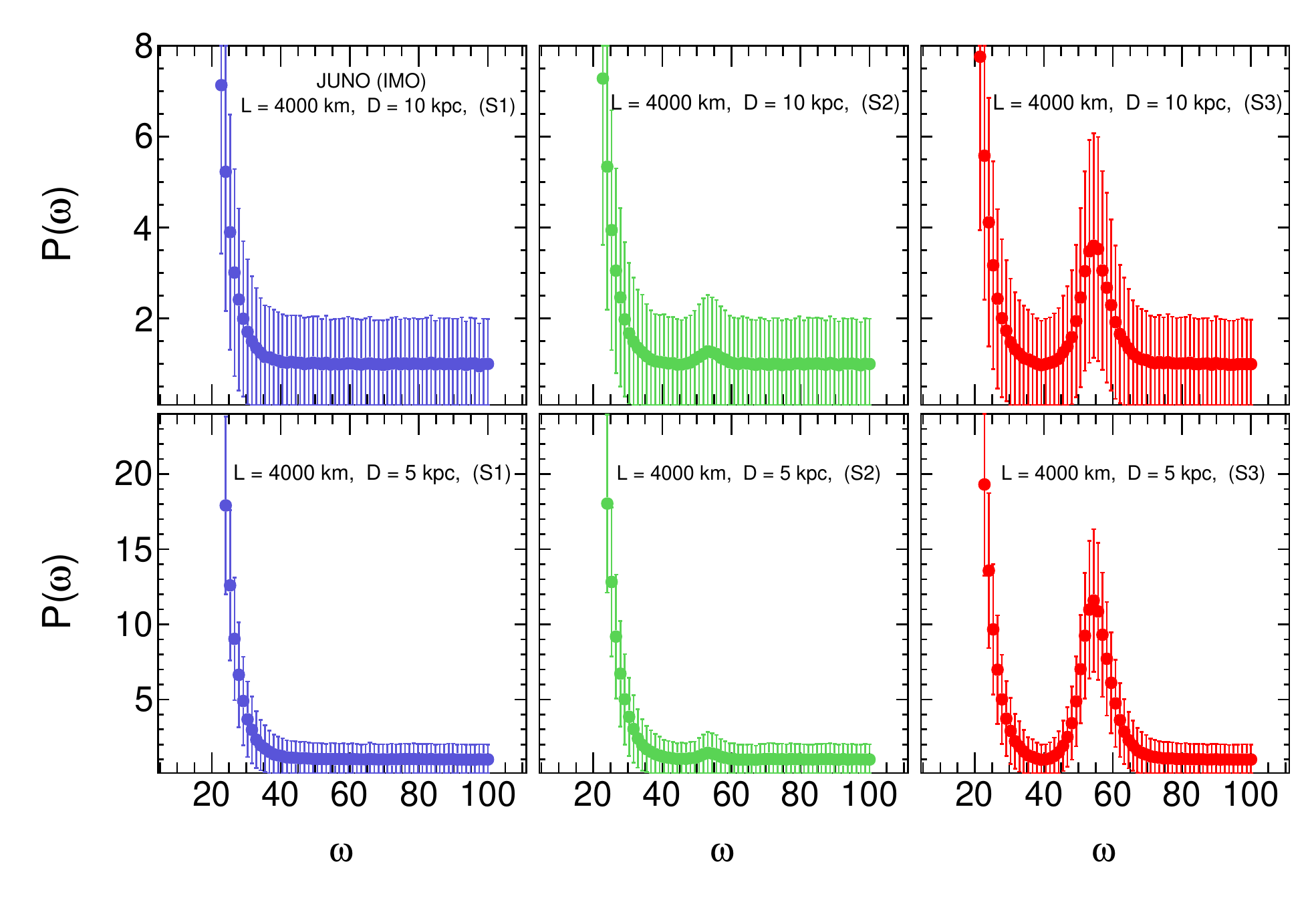}
\vglue -0.5cm
\caption{
 Averaged power spectrum over 1000 MC (Monte Carlo) run samples for $L = 4000$ km  and inverted mass ordering.  We present these results for a CCSN distance of 10 kpc (upper panels) and 5 kpc (lower panels)  for the 3 scenarios, S1 (no decay), S2 ($\sim$ 50\% decay) and S3 (100\% decay) described in the subsection~\ref{subsec:nu-decay-scenarios}.  As error in the measurement, the standard deviation given in the fit has been considered and indicated by vertical error bars. }
\label{figure:pow_spectrum_juno_gg_4000km_IO}
\end{center}
\vglue -0.5cm
\end{figure}

 In the short duration of CCSN neutrino burst, $\sim$ 10 s, IBD reactions  are the main detection channel for $\bar{\nu}_e$ events.  For simplicity,  in this work the only errors to be taken into account are  statistical ones ($1\sigma$ Poisson fluctuations  at the event number determination) and ignore any systematic uncertainties of the detectors.  The propagation of these errors to the power spectrum make difficult to observe the Earth matter effect peaks. With the aim of visualizing how these uncertainties work on the power spectrum,  in figures~\ref{figure:pow_spectrum_juno_gg_4000km_NO} for NMO and ~\ref{figure:pow_spectrum_juno_gg_4000km_IO}  for IMO  we present an example  in which  to obtain each point, a total of 1000 MC (Monte Carlo) run samples have been fitted.  As error in the measurement, the standard deviation given in the fit has been taken (vertical bars).  The left panels refer to S1 scenario,  the central panels to S2 and the right ones to S3. As we can see from the S1 scenario  in figure~\ref{figure:pow_spectrum_juno_gg_4000km_IO}, the expected value for the background fluctuations is $\sim 1$;  this makes that at $1\sigma$  C.L.  the scenarios S1 (for NMO) and S2 (for both NMO and IMO)  are not suitable for Earth matter effects observation, since at less  $68\%$ of observed CCSNe (samples) would present a peak in the power spectrum comparable to or shadowed by the background statistical fluctuations, even for a more optimistic case, such as a CCSN at 5 kpc.

This makes us to focus more on the scenario S3,  which seems to present a greater possibility of identifying the presence of Earth matter effects in the neutrino spectrum and, therefore, be able to observe a clear peak.
  
\begin{figure}
\begin{center}
\includegraphics[scale= 0.65]{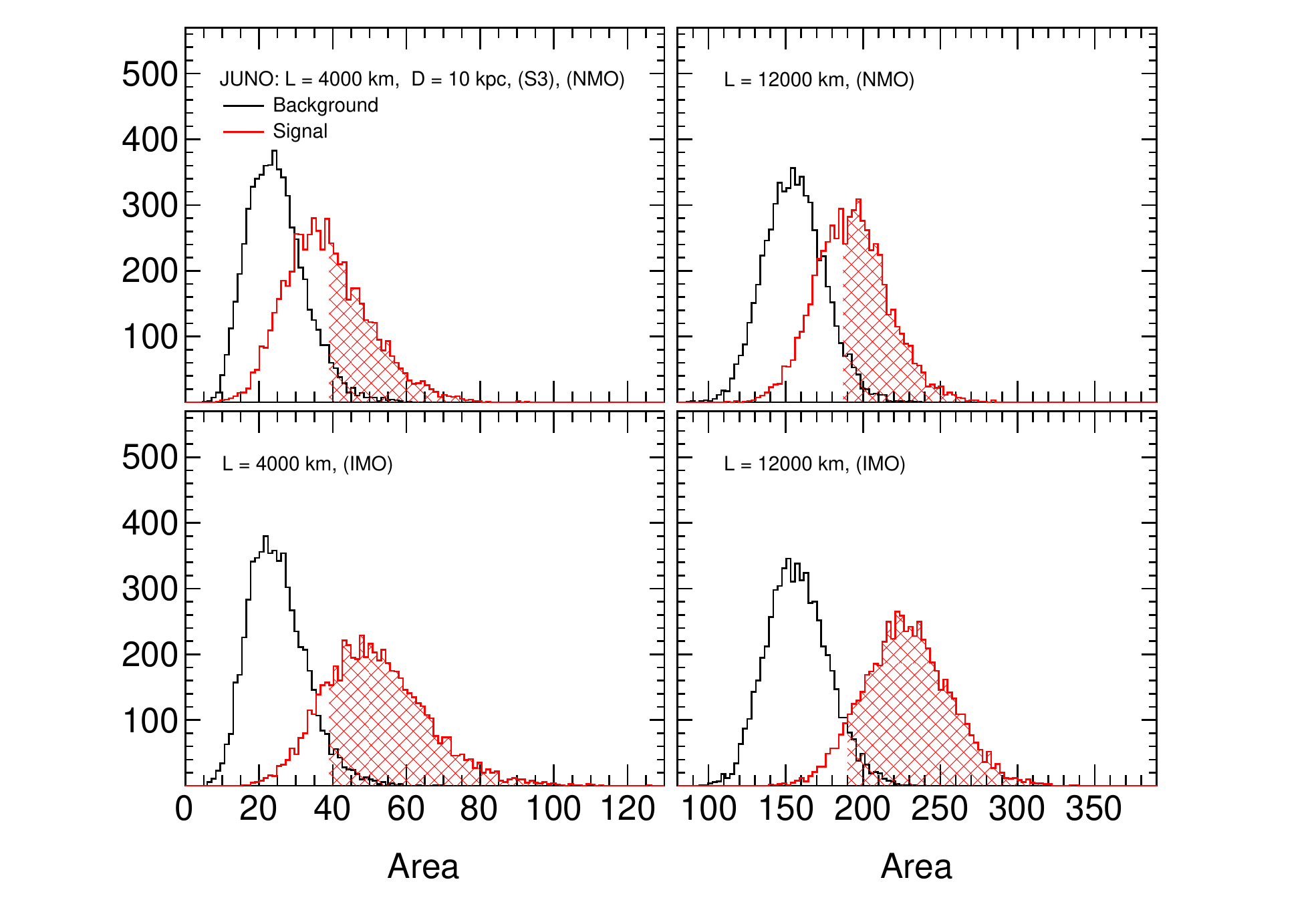}
\vglue -0.5cm
\caption{
Area distribution of the background fluctuations (black) corresponding to the case in the absence  of  the Earth matter effect and that of the signal (red) for  the case in the presence of the Earth matter effect for JUNO are shown for the normal mass ordering in the upper panels
and the inverted mass ordering in the lower panels.  The hatched region correspond to $2\sigma$ (95.45\% C.L.). }
\label{figure:area_distr}
\end{center}
\vglue -0.5cm
\end{figure}

In order to set a benchmark on the Earth effects observation probability for the next CCSN neutrino spectrum, we opted for using the general framework provided by {\it frequentist statistics}  to study decisions that are made in uncertain or ambiguous situations. First, for each sample we calculate the area ($A$) under the curves of  the averaged power spectrum shown in figures~\ref{figure:pow_spectrum_juno_gg_4000km_NO} and ~\ref{figure:pow_spectrum_juno_gg_4000km_IO}; this calculation is made between two fixed frequencies $\omega_{min}$ and $\omega_{max}$  as performed in ref.~\cite{Dighe:2003vm}. Thus we obtain two distributions which we call {\it signal}, corresponding to the case in which we consider the regeneration factors for both S1 and S3, and {\it background} calculated without taking into account the regeneration factors.  Once these distributions are known, we decided to accept the observation of Earth effects at confidence level (C.L.) $1 - \alpha$ if the observation, $A$,  is greater than a critical value $A_c^{\alpha} $ which is known as detection condition. Figure~\ref{figure:area_distr} shows these area distributions for neutrinos coming from a galactic CCSN for  two baselines: $L = 4000$ km  and $L = 12000$ km.

Then, as in ref.~\cite{Dighe:2003vm} we define  the probability of detection, $p = 1 - \beta$, as the fraction of the area of the signal distribution, $p(A\,|\,signal)$,  above $A_c^{\alpha} $; red shaded region in figure~\ref{figure:area_distr}. Where $\beta$ is the probability of making an error type-II, i.e., considering as background fluctuations the signal below the detection criterion.

Unfortunately, this hit probability in the peak identification is incomplete as to indicate how well the Earth effects are detected. There is the possibility of confirming an observation of the peak in question, even though only background fluctuations are present; fraction of the area of the background distribution above the decision criterion  in figure~\ref{figure:area_distr}. This probability is known as probability of false alarm, $\alpha$, or  type-I error rate,  which is important when the two area distributions  considerably overlap.  Usually, this probability is converted into a number of double-sided Gaussian standard deviations. In our case, we will accept a clear Earth effects detection when signal data are more than $n\sigma$ away from the background mean. The conversion between $n\sigma$ and $\alpha$ is given by
\begin{equation}
\alpha(n) = \frac{2}{\sqrt{2\pi}} \int_{n}^{\infty} dx\,e^{-x^{2}/2} = \rm{erfc}\left(\frac{n}{\sqrt{2}}\right)\,,
\label{eqn4_1}
\end{equation}
where erfc(x) is the complementary error function. This definition implies that we identify $1\sigma$, $2\sigma$ and $3\sigma$ with a C.L. $(1 - \alpha)$ of 68.27\%, 95.45\% and 99.73\%, respectively~\cite{An:2015jdp,Blennow:2013oma,Ciuffoli:2013rza}.

Apart from evaluating $\alpha$, it is important to look for more statistical tools that help us to have a better idea of how feasible it is to observe Earth matter effects at a certain level of confidence; one of these tools is the sensitivity  to discriminate between the signal and the background fluctuations which can be  defined as the separation between the means of the signal and the background fluctuation distributions, compared against the standard deviation of the signal and background fluctuation distribution.  Mathematically it can approximately be written as
\begin{equation} \label{eqn4_6}
d'  = \frac{\mu_{s} - \mu_{b}}{\sqrt{\frac{1}{2}\left(\sigma_{s}^2 + \sigma_{b}^2\right)}}\,, 
\end{equation}
\noindent where $\mu_{s}$ and $\mu_{b}$ are the means of the signal and background fluctuation distributions  and $\sigma_{s}$ and $\sigma_{b}$ their standard deviations, respectively. As we can see, when we compare the  left and  right panels in figure~\ref{figure:area_distr}, this statistic has a higher value when neutrinos cross the Earth's core on their way to the detector than when they cross only the mantle, and if we compare the  upper and lower  panels we can observe that inverted mass ordering offers a better sensitivity to discriminate between signal and background fluctuations.

\begin{figure}[H]
\begin{center}
\includegraphics[scale=0.65]{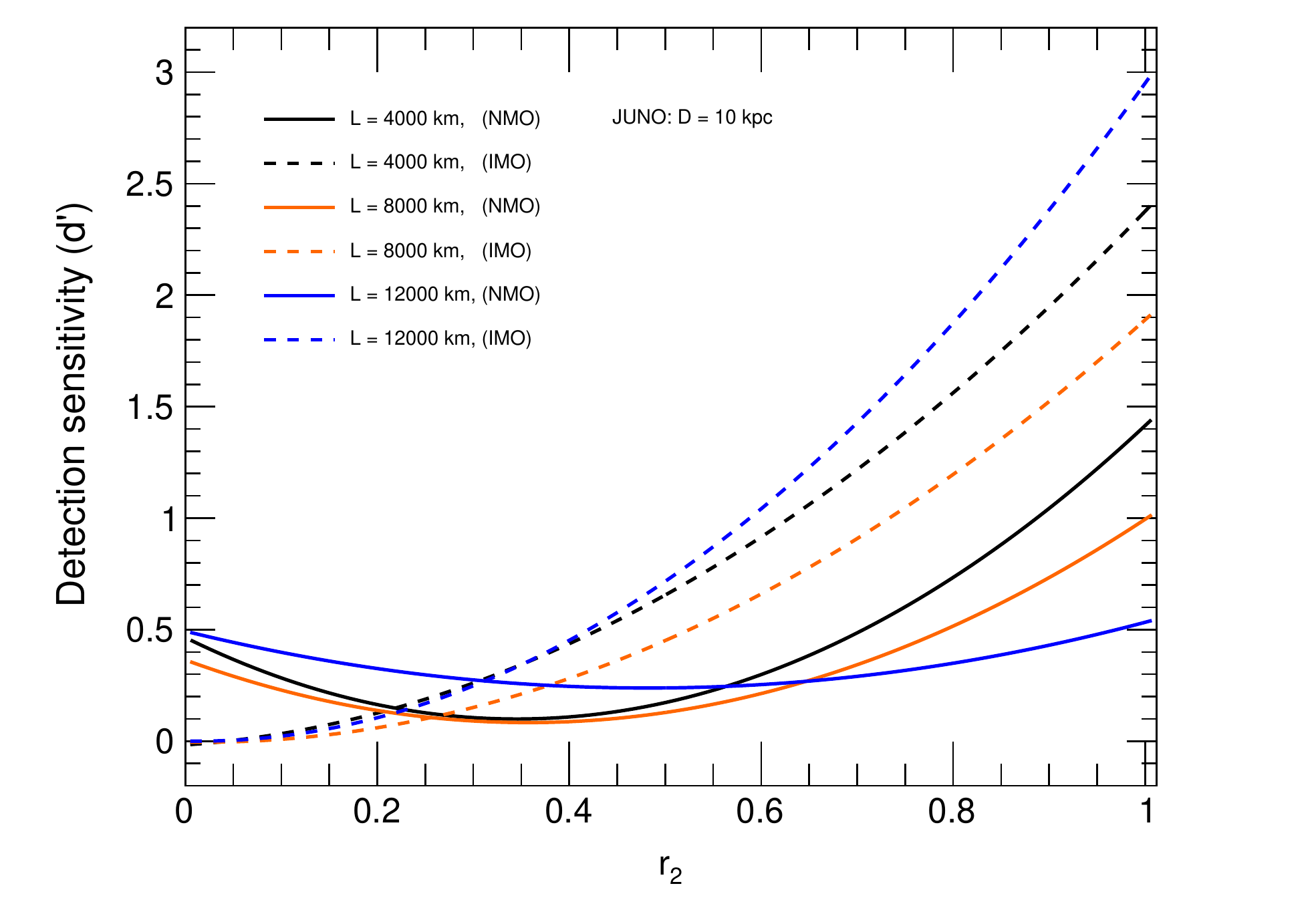}
\vglue -0.5cm
\caption{
Sensitivity to Earth matter effects detection in presence of invisible neutrino decay for JUNO as a function of the $\nu_2$ decay rate $r_2$.}
\label{figure:sensitivity_juno}
\end{center}
\vglue -0.5cm
 \end{figure}

Figure~\ref{figure:sensitivity_juno} shows another characteristic of $d'$, its dependence on $r_{2}$, the decay rate of $\nu_2$.  Only for the results shown in figure~\ref{figure:sensitivity_juno}  (and also in  figure~\ref{figure:sensitivity_dune}), for simplicity, we compute CCSN neutrino fluxes by setting $r_2$ as energy independent constant in eqs.~\eqref{eq:flux_NO_decay} and ~\eqref{eq:flux_IO_decay},  unlike the one shown in eq.~\eqref{eq:r-i}. We observe that for the case of NMO (solid curves), as $r_2$ increases, the detection sensitivity starts to decrease in the beginning and then take some minimum values around $r_2 \sim  0.4$ for $L$ = 4000 km or $L$ = 8000 km and $r_2 \sim  0.6$  for $L$ = 12000 km,  and then increase up to $r_2=1$. On the other hand, for the IMO case, the sensitivity (dashed curves) increase monotonically up to $r_2=1$.

We can try to understand qualitatively such a difference of behaviors of the sensitivity between NMO and IMO found in figure~\ref{figure:sensitivity_juno} as follows.  We observe in the lower left panels of figures~\ref{figure:Delta_F_4000km}-\ref{figure:Delta_F_12000km} corresponding to NMO  for $\bar{\nu}_e$,  for the scenario S1 or $r_2 = 0$ (no decay), the $\Delta F_{\bar{\nu}_e}$ is negative in the relevant neutrino energy range of $\gtrsim 15$ MeV  whereas for the scenarios of larger decay rates, S2 and S3,   $\Delta F_{\bar{\nu}_e}$ are mostly positive for all the energy range.  This implies that when the decay rate increases from 0 to 100\%,  $\Delta F_{\bar{\nu}_e}$
($=  \bar{f}_\text{reg}[(1-r_2)F_{\bar{\nu}_x}^0-F_{\bar{\nu}_e}^0 ]$)  around  the relevant energy range changes its sign.   At the decay rate corresponding to the point where $\Delta F_{\bar{\nu}_e}$ changes its sign,   $\Delta F_{\bar{\nu}_e} \sim 0$ around the energy of $\sim 15-25$ MeV which correspond to the peak  of the events (see figure~\ref{figure:event_distr_juno_gg_10kpc_4000km}),  and therefore, we expect worst Earth matter effect detection sensitivity around such a point.  On the other hand, if we look at the lower right panels of 
figures~\ref{figure:Delta_F_4000km}-\ref{figure:Delta_F_12000km} corresponding to IMO,  we can see that $\Delta F_{\bar{\nu}_e}$ are always (or mostly) positive for all the energy range,  increasing monotonically as  the decay rate increases.

We can also try to understand the behaviors of the curves for NMO in figure~\ref{figure:sensitivity_juno}  based on the discussions given in \cite{Borriello:2012zc}  in which it is argued that a suppression in the intensity of the peaks is obtained  due to partial destructive interference presented by the different components of  the power spectrum (i.e. $P_{\bar{\nu}_e}(\omega)$,  $P_{\bar{\nu}_x}(\omega)$ and $P_{\bar{\nu}_e, \bar{\nu}_x}(\omega)$).  For $r_2$ larger than 0 but less than $\sim$ 0.4-0.6 (depending on the value of $L$), due to the decay effect, the effective ratio of the neutrino fluxes $\Phi^0_{\bar{\nu}_e}/\Phi^0_{\bar{\nu}_x}$ increases which leads to the destructive interference between the different components of the power spectrum~\cite{Borriello:2012zc} exhibiting ``double peak'' structures as we can seen in the left panel of figure 2 of \cite{Borriello:2012zc} as well as in lower middle panel of our figure~\ref{figure:pow_spectrum_juno_gg_4000km_NO}.

On the other hand, when $r_2$ goes up to 1 from the values of $\sim$ 0.4-0.6, the power spectrum presents a single-peak-like structure whose intensity begins to increase monotonically as the $\bar{\nu}_2$ decay increases like the peaks we can seen in the case of S3 in
figures~\ref{figure:pow_spectrum_juno_gg_4000km_NO} and \ref{figure:pow_spectrum_juno_gg_4000km_IO}.  Therefore, sensitivity to Earth effects detection begins to increase. Thus, figure~\ref{figure:sensitivity_juno} can help us to understand why decay scenarios such as the scenario S2  for NMO ($< 70 \%$  of $\bar{\nu}_2$ decay)  does not enhance the impact on Earth effect identification  when compared to S1.

Hereafter, in this paper,  we will focus on the scenario S3, whose configuration presents a more optimistic scenario for an unambiguous identification of the Earth matter effects via Fourier transform of the inverse-energy spectrum.  As we have already pointed out, a clear peak identification requires  a good energy resolution of the detector as well as the observation of a large number of events. Since the main goal of JUNO  is to establish the neutrino mass ordering by analyzing the  $\bar{\nu}_e$ IBD spectrum, the energy resolution that  this detector must achieve is  $\leq 3\%$ at the visible energy $\sim 1 $ MeV. Assuming such a energy resolution, we proceed to study the behavior of the detection probability, $p$, with respect to CCSN distance for the following 3 neutrino emission model assumptions:
Model A, with our default choice of CCSN parameters mentioned in the end of the subsection~\ref{subsec2_1}  similar to the ones  considered in \cite{Borriello:2012zc},  model B, the numerical results obtained by the Lawrence Livermore (LL) group ~\cite{Totani:1997vj}, and  model C, results of the Monte Carlo study of spectral formation by Keil, Raffelt, \& Janka (KRJ accretion phase model II)~\cite{Keil:2002in}.  In table~\ref{table1}, we give the information on the value of the parameters used in these models, where we have also assumed that  $L_{\nu_e} = L_{\bar{\nu}_e}  = L_{\nu_x} = 5 \times 10^{52}$ ergs.   Since $\beta_{\nu_e}$ is not given in the LL model, the same value as $\beta_{\bar{\nu}_e}$ has been assumed~\cite{Lunardini:2016}.
\begin{table}
\centering
\begin{tabular}{|c|c|c|c|c|c|c|}\hline
Model &  $\left< E_{{\nu}_e} \right> $ &  $\left< E_{\bar{\nu}_e} \right> $ & $\left< E_{\bar{\nu}_x} \right> $ & $\beta_{\nu_e}$ & $\beta_{\bar{\nu}_e}$ & $\beta_{\nu_x}$ \\\hline
A   &  13        &  15         &   18         &   4          &   4        &  4         \\
B   &  11.2     &  15.4      &   21.6      &   4.8       &   4.8     &  2.8      \\
C   &  13.0     &  15.4      &   15.7      &   4.4       &   5.2     &  3.5      \\\hline
\end{tabular}
\caption{
CCSN neutrino emission model parameters used in this work.  The values for the model B (LL) were taken from table 1 of the the first version of the preprint of ref.~\cite{Lunardini:2016} whereas that for the model C (KRJ II) were taken from \cite{Keil:2002in}. 
}
\label{table1}
\end{table}

\begin{figure}[H]
\begin{center}
\includegraphics[scale=0.65]{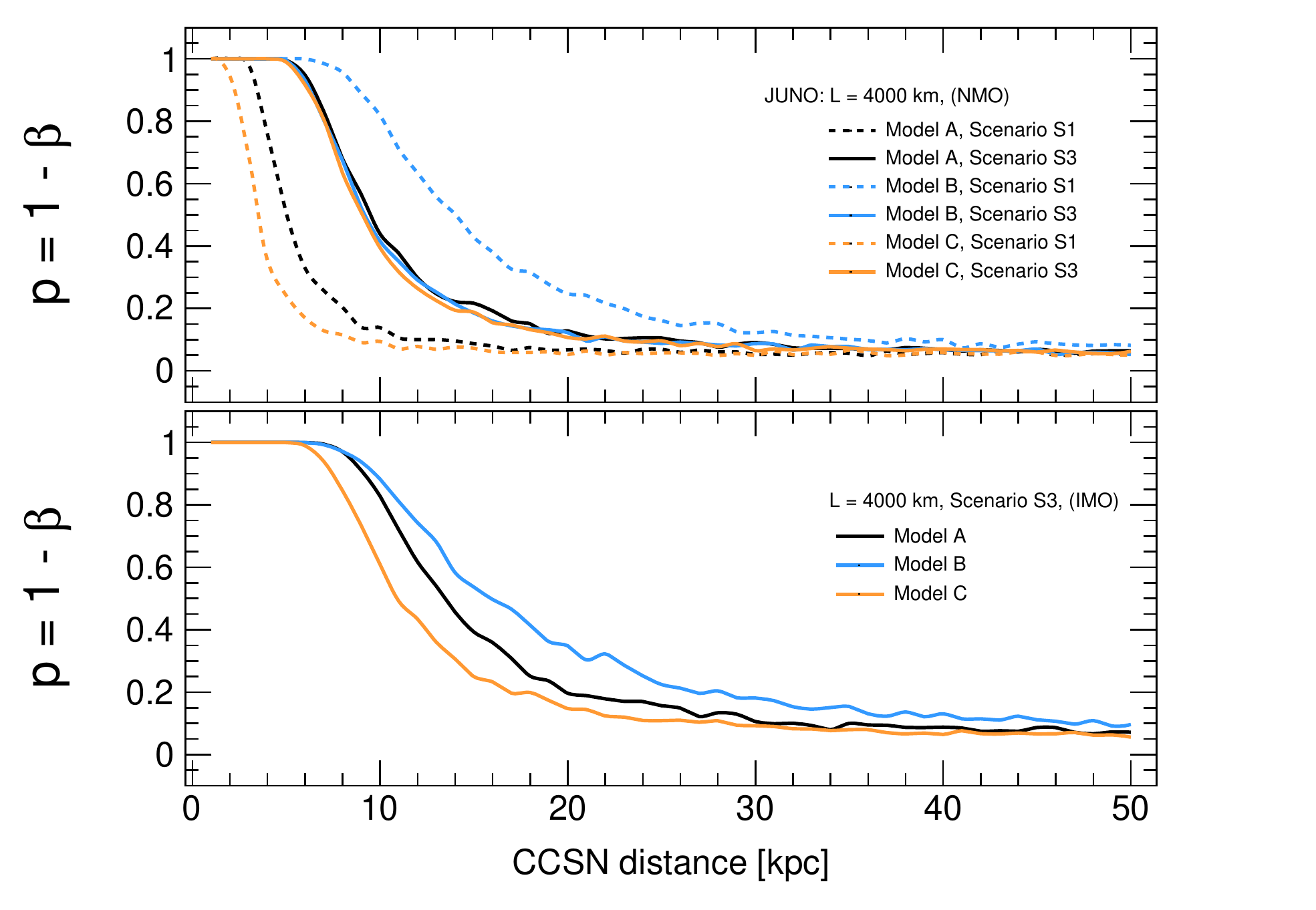}
\vglue -0.5cm
\caption{
Probability that Earth matter effects can be observed at $2\sigma$ (95.45\% C.L.) by JUNO for NMO (upper panel)  and IMO (lower panel) for different CCSN  emission models A, B (LL) and C (KRJ II) as a function of the distance to CCSN.}
\label{figure:id_prob_juno}
\end{center}
\vglue -0.5cm
\end{figure}

Figure~\ref{figure:id_prob_juno} shows for NMO (upper panel) and IMO (lower panel),  the probabilities of successfully identifying a peak
(in power spectrum) associated with Earth matter effects in a range of distances that would include stars such as Betelgeuse, Mira Ceti and Antares, that in the future could explode as CCSN, as well as possible candidates from the Large Magellanic Cloud, a dwarf satellite galaxy of the Milky Way.

We first observe that for the NMO, all the 3 CCSN models lead to the similar results (see solid curves) for the scenario S3 (100\% decay).
  We believe that this is because for this case,  $\Delta F_{\bar{\nu}_e} = -\bar{f}_{\text{reg}} F^0_{\bar{\nu}_e}$ (see eq. \eqref{eq:Delta_F_Earth_NO_IO}) are expected to be similar for 3 models considered as CCSN parameters for ${\bar{\nu}_e}$ are not very different, especially the values of $\left< E_{\bar{\nu}_e} \right> $.  On the other hand, if we compare the scenarios S3 and S1 (no decay), with the decay effect,  the models A and C lead to stronger Earth matter effects (leading to stronger modulations of the spectrum) 
  while the model B lead to the opposite, or no decay leads to larger Earth matter effect.  We believe that this is because the model B in the absence of decay effect, exhibits large difference between ${\bar{\nu}_e}$  and ${\bar{\nu}_x}$  spectra, leading to the larger Earth matter effect, as $\Delta F_{\bar{\nu}_e} = -\bar{f}_{\text{reg}} (F^0_{\bar{\nu}_e}-F^0_{\bar{\nu}_x})$.

For IMO (in the lower panel of figure~\ref{figure:id_prob_juno}),  we show only the results for the scenario S3 since there is no Earth matter effect for S1 (no decay) for $\bar{\nu}_e$ \cite{Dighe:1999bi}.  For this case the results for the 3 models are not very similar.  We believe that this is because  $\Delta F_{\bar{\nu}_e} = -f_{\text{reg}} F^0_{\bar{\nu}_x}$ for this case, and the CCSN parameters for these 3 models for $\bar{\nu}_x$ are  not so similar, see table \ref{table1}.

Here,  we can see that distances $\leq 6$ kpc  have a good chance, $p \sim 1$, while distances $\geq 40$ kpc have a zero or very small chance in practice.  In particular, for a typical CCSN at region close to the galactic center ($\sim 10$ kpc)  we find  for the model A  a  detection probability  of 44\% (81\%) for NMO (IMO) compared to $\sim 5\%$  of false alarm probability; detection probability is only 12\%  at the same C.L. in absence of neutrino decay (i.e., S1).

The main characteristics (mentioned above) observed in figure~\ref{figure:id_prob_juno} remain for $L$ = 8000 km and $L$ = 12000 km, finding that the JUNO resolution allows it to verify for NMO that detection probability is greater when neutrinos cross the core of the earth (three peaks are identified in the power spectrum) than when only the mantle is crossed. 
 
At this point, it is important to note that the probability of identifying Earth matter effects depends on the used detection criterion: a higher one (e.g., $3\sigma$ C.L.) produces a better background rejection but few hit to discriminate signal from the background fluctuations, while a low detection criterion (e.g., $1\sigma$ C.L.) increased the hit rate at the cost of the correct rejection rate.
\begin{figure}[H]
\vglue -0.5cm
\begin{center}
\includegraphics[scale=0.65]{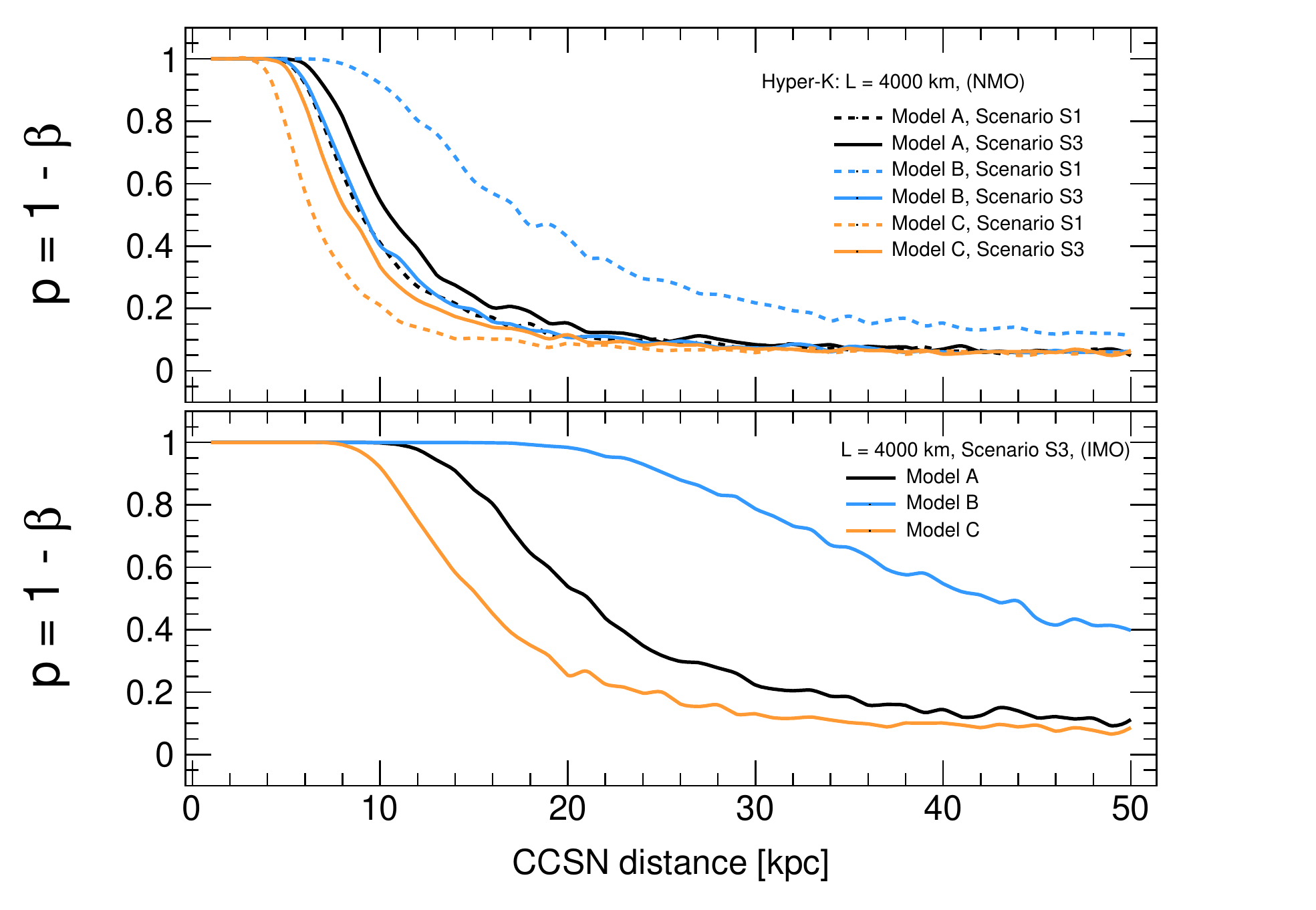}
\vglue -0.5cm
\caption{
Probability that Earth matter effects can be observed at $2\sigma$ (95.45\% C.L.) by Hyper-Kamiokande for NMO (upper panel) and IMO (lower panel) for different CCSN  emission models A, B (LL) and C (KRJ II) as a function of the distance to CCSN.}
\label{figure:id_prob_hk}
\end{center}
\vglue -0.5cm
\end{figure}
%


\subsection{The decay impact for Hyper-Kamiokande}
\label{subsec-HK}
Next let us discuss the decay impact for the Hyper-Kamiokande(-like) detector. Hyper-Kamiokande (or Hyper-K)~\cite{Abe:2018uyc} is the next generation Water Cherenkov  detector to be located  in Tochibora mine at 295 km away from the J-PARC proton accelerator research complex in Tokai, Japan. The detector has the size of 60 m height and 74 m in diameter with 40\% photo-coverage~\cite{Abe:2018uyc}.
The construction of the detector has been started in 2020 while the data taking is expected to start in 2027.  The major goals of HK is to observe CP violation in the lepton sector in the long-baseline neutrino oscillation   as well as to search for nucleon decay. In addition, it is important for HK to perform various astrophysics programs including the observation of neutrinos from nearby CCSNe.

Hyper-K has a possibility to have its second detector in South Korea with the same size as the first one which is under construction
~\cite{Hyper-Kamiokande:2016srs} but this is still under discussion. For the first detector in Japan,   we compute the number of events in a similar way as in the case of JUNO by replacing the experimental parameters such as fiducial volume and energy  resolution by that of Hyper-K.  As we have done for JUNO, we consider only the main features of the Hyper-K detector, namely, size and its energy resolution,
to approximately simulate it.  For CCSN neutrino analysis, we can consider 220 kt ~\cite{Abe:2018uyc} which implies that the number of target (free protons) in the detector is $N_T \sim 1.47\times10^{34}$.  The main detection mode for CCSN neutrinos is the IBD reaction. The energy resolution of the Hyper-Kamiokande detector is expected to be $\sim \sqrt{2} $ times better than that for Super-Kamiokande (SK) detector assuming the same photo-coverage as SK but with $\sim 2$ times better sensitivity of PMTs. Since SK's energy resolution is 14.2\% at $E_\nu$ = 10 MeV~\cite{Suzuki:2019jby}. Therefore, in this work we assume the similar Gaussian type energy resolution function given in \eqref{eq:energy-resolution-function} but with $\delta E_v$/MeV = 32\% $\sqrt{E_\nu/\text{MeV}}$. The expected number of IBD events is about 50,000 to 75,000 for a CCSN at 10 kpc from the Earth~\cite{Abe:2018uyc}.

In figure~\ref{figure:id_prob_hk}  we can see that Hyper-K presents good opportunities to detect Earth matter effects, even better than JUNO for $L$ = 4000 km. However, it is worth noting that its lower energy resolution prevents a good identification (NMO scenario) of Earth effects for the baselines $L$ = 8000 km and $L$ = 12000 km for CCSN distances greater than 5 kpc. This is because, for $L=4000$ km, the modulation  in the neutrino energy spectrum is slow (as a function of the inverse of neutrino energy) such that it is easier to detect the modulation compared to the larger $L$ which tends to washout the effect due to faster modulation. 

On the other hand, we observe that most of the characteristics discussed for figure~\ref{figure:id_prob_juno} are preserved, with the notable exception that for IMO the range in which $p \sim 1$ is shifted a little further to the right and in the case of LL model there is a chance of 40\% (as opposed to $\sim 5\%$ for false alarm probability) of Earth effects identification for a CCSN at 50 kpc from us.


\subsection{The decay impact for DUNE}
Finally let us discuss the decay impact for the DUNE(-like) detector. The Deep Underground Neutrino Experiment (DUNE)~\cite{Abi:2020evt} is a future neutrino observatory and nucleon decay detector designed primarily to study neutrino mass ordering and CP violation in the lepton sector as well as to detect and measure the  $\nu_e$ flux from the next galactic CCSN. This experiment will consist of a far detector placed at the Sanford Underground Research Facility (South Dakota, USA) and a near detector to be located at Fermilab. The far detector will be  composed of four liquid argon time-projection chambers (LArTPCs) of 10 kt fiducial mass each and will bring unique electron neutrino sensitivity  to the observation of the  CCSN neutrino burst  via the charged current  process (dominant interaction).
\begin{equation}
\label{eq:nue_Ar_process} 
\nu_e + ^{40}\,Ar \to ^{40}\,K^{*} + e^{-}.
\end{equation}

The expected number of events by this reaction in the DUNE detector is typically $\sim$ 3000 for a CCSN at 10 kpc located away from the Earth~\cite{DUNE:2020zfm}.  As before we consider only the main features of DUNE, size and energy resolution, to make an approximate simulation of the DUNE detector.

In the LArTPCs, the interaction of electron neutrinos in range $\sim$ 5 MeV (threshold energy) to few tens of MeV create short electron tracks which are generally  accompanied  by gamma rays and other secondary particle signatures \cite{Abi:2020lpk}. In this work we use the cross section for the most relevant CCSN neutrino interaction  in argon  found in the  SNOwGLoBES software package \cite{SNOwGLoBES} and the energy resolution that  has been calculated by the ICARUS collaboration~\cite{GilBotella:2003sz}
$\delta E_{v}/\text{MeV}  =   0.11\,\sqrt{E_\nu/\text{MeV}} + 0.02(E/\text{MeV})$ which is better than the one that Hyper-K would have. 
 \begin{figure}
\begin{center}
\includegraphics[scale=0.65]{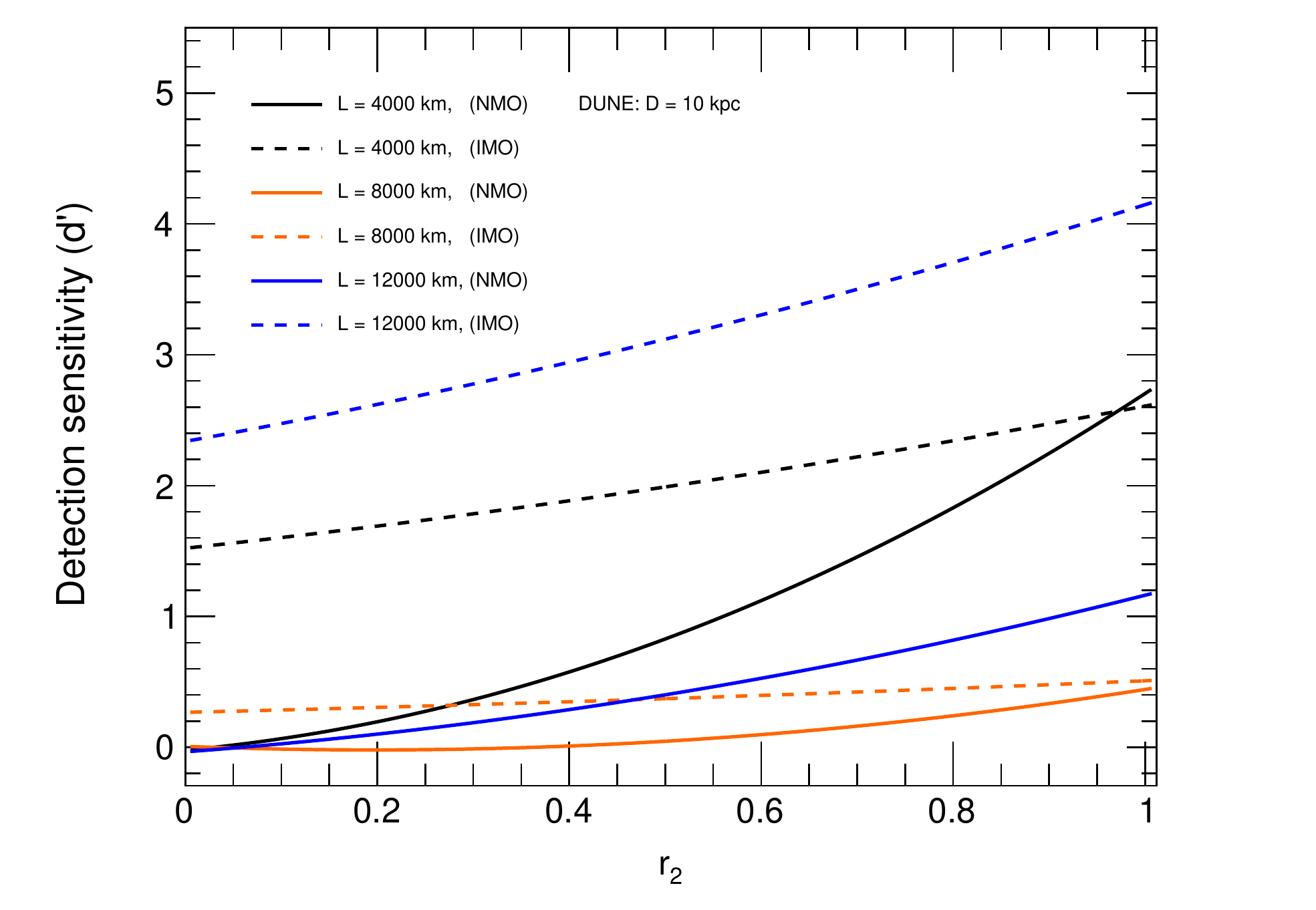}
\vglue -0.5cm
\caption{
Sensitivity to Earth matter effects detection in presence of invisible neutrino decay for DUNE  as a function of the $\nu_2$ decay rate $r_2$.}
\label{figure:sensitivity_dune}
\end{center}
\vglue -0.5cm
\end{figure}

Figure~\ref{figure:sensitivity_dune} shows the DUNE sensitivity,  as a function of $r_2$,  to discriminate between signal and background fluctuations for $\nu_e$ fluxes that crossed the Earth with a baseline $L$ = 4000 km.  As done in figure~\ref{figure:sensitivity_juno}, $r_2$ was set to be an energy independent decay rate for simplicity.  First of all, we note that all curves are {monotonically increasing functions of $r_2$. This behavior seem to be consistent with the results shown in figures~\ref{figure:Delta_F_4000km}-\ref{figure:Delta_F_12000km}.
If we look at the results for $\nu_e$ channel in the upper panels  of these figures, as $r_2$ varied from 0 to 1, apart from the sign change in the lower energy range $\lesssim$ 15 MeV for IMO, the impact of the decay increases. In particular, for NMO, we can also understand analytically that the Earth matter effect should be an increasing function of $r_2$ as $\Delta F_{\nu_e} \simeq  -f_{\text{reg}} r_2 F_{\nu_x}^0$ (see eq. \eqref{eq:Delta_F_Earth_NO}).  Therefore, the ratio between these un-oscillated fluxes is such that it does not lead to
double-peak-like structures formation  (destructive interference) in the power spectrum discussed in \cite{Borriello:2012zc} for all neutrino decay scenarios. In other words, neutrino decay at the electron neutrino channel leads always to a better identification of Earth matter effects than in the absence of neutrino decay.

 Figure~\ref{figure:id_prob_dune} makes visible the capabilities of DUNE to identify Earth matter effects at 2$\sigma$ C.L.  for $L$ = 4000 km.  As we can see, both mass orderings  have a good chance of identifying clear peaks at the power spectrum when the CCSN distance is less than  $\sim 15$ kpc and in the case of the model B also exist a 27\% of probability  to obtain an Earth effect signal  at $2\sigma$ C.L. for a CCSN at 50 kpc. As with JUNO and Hyper-K, we also study the DUNE identification probability for  $L$ = 8000 km and $L$ = 12000 km; for the first distance  we find that DUNE has no chance to detect Earth matter  effects from the next galactic CCSN while for the second there is very good chance for both mass orderings.
\begin{figure}
\begin{center}
\includegraphics[scale=0.65]{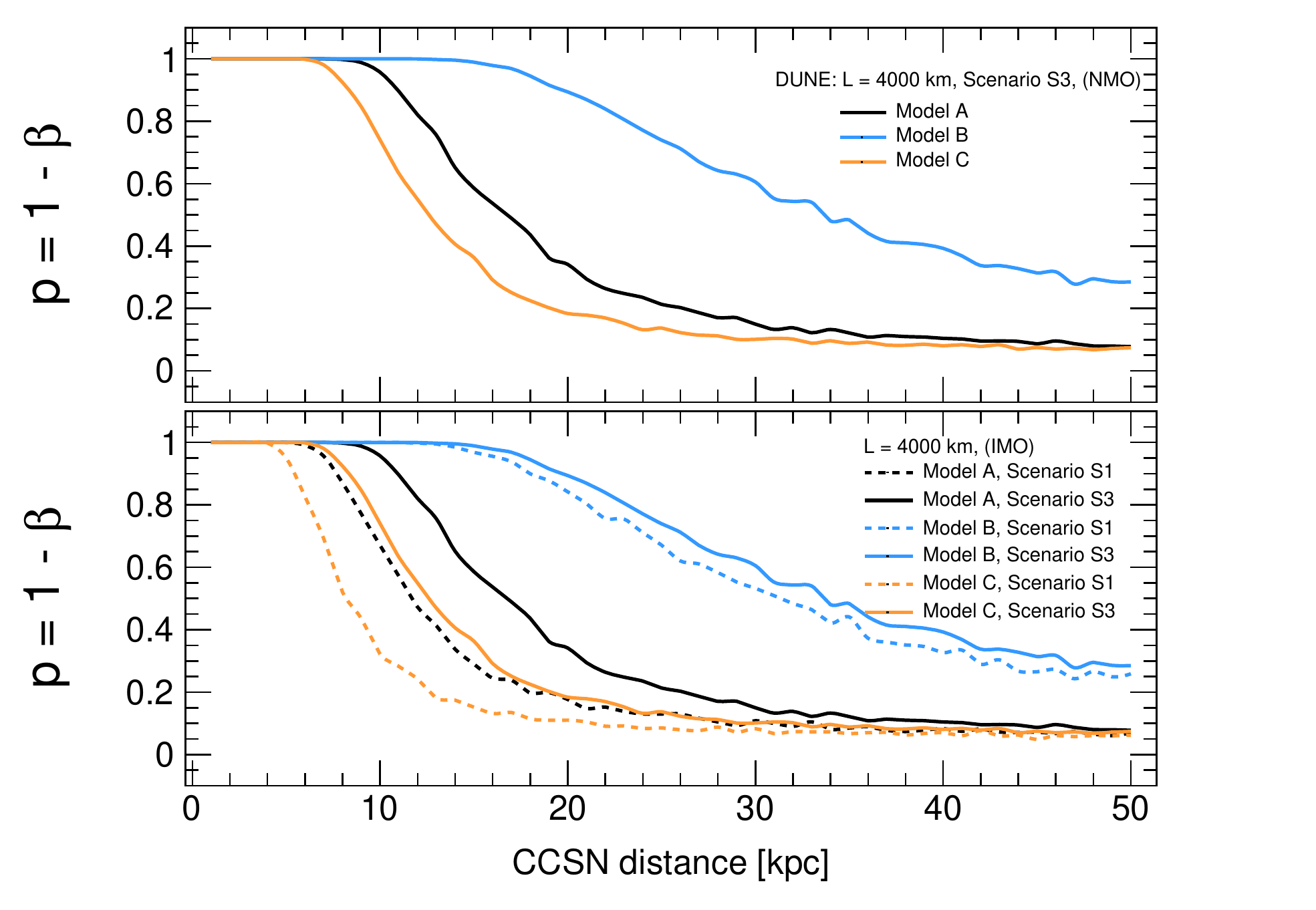}
\vglue -0.5cm
\caption{
Probability that Earth matter effects can be observed at $2\sigma$ (95.45\% C.L.) by DUNE for NMO (upper panel) and IMO (lower panel) for different CCSN  emission models, A, B (LL) and C (KRJ II) as a function of the distance to CCSN.}
\label{figure:id_prob_dune}
\end{center}
\vglue -0.5cm
\end{figure}
%


\section{Summary and conclusions}

In summary, we have focused on the impact of neutrino decay on the possibility of detecting Earth matter effects using a single detector.
Adopting a traditional  supernova emission model as an example and carrying out a Fourier analysis of the inverse-energy spectrum of the CCSN neutrino signal, we show the positive impact that neutrino decay has on the identification of Earth matter effects if $\nu_2$ decay rate is greater than $\sim$ 70\%.  

In this work, we have considered, as our default future detectors, the ones which have the main features of JUNO, Hyper-K and DUNE detectors and we have studied three $\bar{\nu}_2/\nu_2$ decay scenarios - S1 (no decay), S2 ($\sim 50\%$) and  S3 ($100\%$ of decay).
We stress that an unambiguous identification of the Earth effects via Fourier transform of the inverse-energy spectrum requires observing a large number of events. This in turn depends on the detector size, its  energy resolution, the CCSN model parameters which are not very well known and the distance to CCSN which may not be well determined.

For both mass orderings, we investigate the possibilities of Earth effects detection for different baselines (4000, 8000 and 12000 km) and CCSN distances,  for the three CCSN neutrino emission models, A (our default choice), B (Lawrence Livermore group)  and C (accretion phase model II of Garching group). Based on these studies, we can establish the following conclusions:

\begin{itemize}
\item[(1)] In the presence of neutrino decay, Earth matter effect can be present for both  $\nu_e$ and $\bar{\nu}_e$ channels at the same time for both mass orderings. Therefore, if the Earth matter effect will be observed for both channels at the same time, this could be an indication of the presence of neutrino decay as this is not expected to occur in the standard scenario. 
\item[(2)] For the $\bar{\nu}_e$ channel, there is a greater possibility of observing Earth effects if mass ordering is inverted.
\item[(3)]
Due to the finite number of CCSN neutrino events that the current and future detectors can observe, the chance to identify a clear Earth effect signal is overshadowed by the presence of background statistical fluctuations. The capability of a particular detector to discriminate between background and signal basically depends on the number of observed events and the detection technology (eg, liquid scintillator, water Cherenkov, liquid argon, etc.) 
\item[(4)] When neutrinos only traverse the mantle, there is a greater possibility of observing Earth effects for small baselines,  such as $L$ = 4000 km. This is because the modulation as a function of neutrino energy is slower compared to longer baselines such that it is easier  to be detected due to the limited energy resolution.
\item[(5)] Depending on the energy resolution of the detector, the passage of neutrinos through the Earth core enhances the peaks observability in the power spectrum associated with the next galactic CCSN. Among our detectors only JUNO has enough resolution to distinguish three peaks for S3.
\item[(6)]
Due to the formation of double-peak-like structures in the $\bar{\nu}_e$ power spectrum for normal mass ordering, decay scenarios such as the scenario S2 does not enhance the impact on Earth effect identification when compared to S1. 
\item[(7)] For a typical galactic CCSN at $10.7 \pm 4.5$ kpc ~\cite{Mirizzi:2006xx}  we find that neutrino decay increases the Earth matter effect detection probability except to the model B (LL) (exhibiting the presence of a double-peak-like structure for almost all decay scenarios). However, considering that recent CCSNe simulations indicate lower average energies during the cooling phase than the ones obtained by LL, we should not worry about the negative impact that such a model presents when  we explore neutrino decay via Earth matter effects. In any case, a decrease in peak observation probability (when a higher value is expected in the absence of decay) would be a sub-dominant manifestation of the presence of neutrino decay at the CCSN spectrum. 
\end{itemize}


\begin{acknowledgments}

This study was financed in part by Conselho Nacional de Desenvolvimento Cient\'{\i}fico e Tecnol\'ogico - CNPq and by the Coordena\c{c}\~ao de Aperfei\c{c}oamento de Pessoal de N\'{\i}vel Superior - Brasil (CAPES) - Finance Code 001.  HN thanks the hospitality of Anatael Cabrera and the IJCLab where the some part of this work was performed.  We also thank Takatomi Yano for useful correspondence regarding the Hyper-Kamiokande detector.

\end{acknowledgments}


\appendix

\section{Peaks in the power spectrum of CCSN neutrinos: miscellaneous cases}
\label{Appendix1}

Experimentally, neutrino signal is observed as a discrete  set of events, so the Fourier Transform must be defined in this context. Following ref.~\cite{Liao:2016uis} , we define the power spectrum of the $N$ detected events as
\begin{equation} \label{eqn4_5}
P(\omega)  = \left|\frac{1}{\sqrt{N}}\sum_{\rm{Energy\, bin\, i}} \frac{N_i}{\Delta y_i} \int_{\Delta y_i} dy_i\, e^{i\omega y_i}\right|^2\,, 
\end{equation}
where $y = 12.5/E$ MeV is the inverse energy, $\Delta y_i$ is the width of the i-th bin and $\omega$ is the frequency that characterizes the  modulations of the inverse energy spectra.

We show the power spectrum for the JUNO detector  for several baselines (columns) and distances from the CCSN (rows) for NMO in figure~\ref{figure:pow_spectrum_misc_NO_JUNO} and IMO in figure~\ref{figure:pow_spectrum_misc_IO_JUNO}.  We also show the similar plots for Hyper-K for NMO in figure~\ref{figure:pow_spectrum_misc_NO_HK}  IMO in figure~\ref{figure:pow_spectrum_misc_IO_HK}, and for DUNE in
figure~\ref{figure:pow_spectrum_misc_NO_DUNE} for NMO and in figure~\ref{figure:pow_spectrum_misc_IO_DUNE} for IMO. In these figures, we considered the distance to CCSN of $D = 1, 5, 10$ and 15 kpc whereas the distance traveled by neutrinos inside Earth as $L$ = 4000, 8000 and 12000 km, and consider the scenario S3 (100\% decay). 

\begin{figure}
\begin{center}
\includegraphics[scale=0.7]{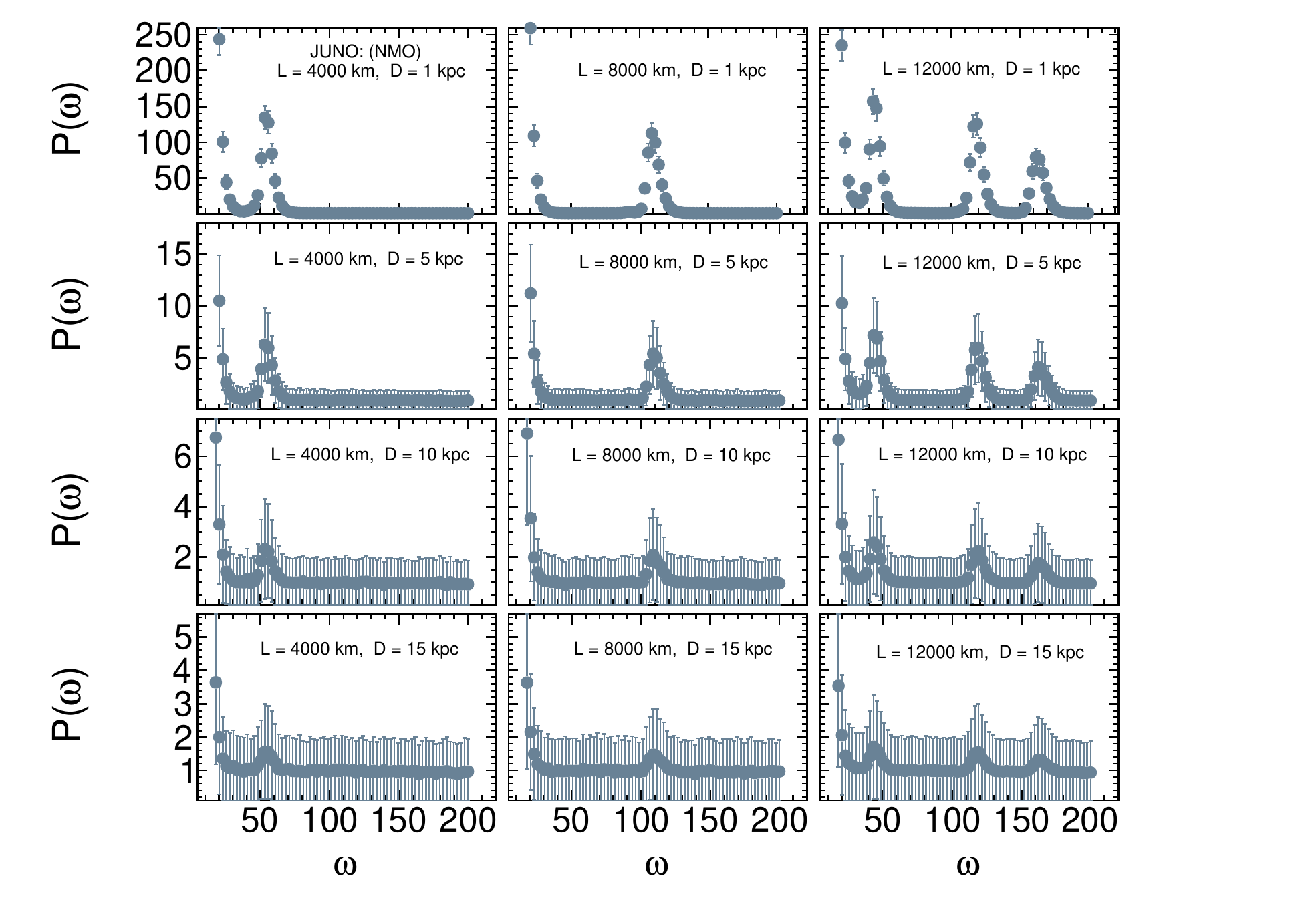}
\vglue -0.4cm
\caption{
CCSN power spectrum at JUNO detector  for several baselines (columns) and distances from the CCSN (rows) for the normal mass ordering.
We have considered only the scenario S3 (100\% decay) for visualizing the different changes.}
\label{figure:pow_spectrum_misc_NO_JUNO}
\end{center}
\vglue -0.5cm
\end{figure}
%
\begin{figure}[H]
\begin{center}
\includegraphics[scale=0.7]{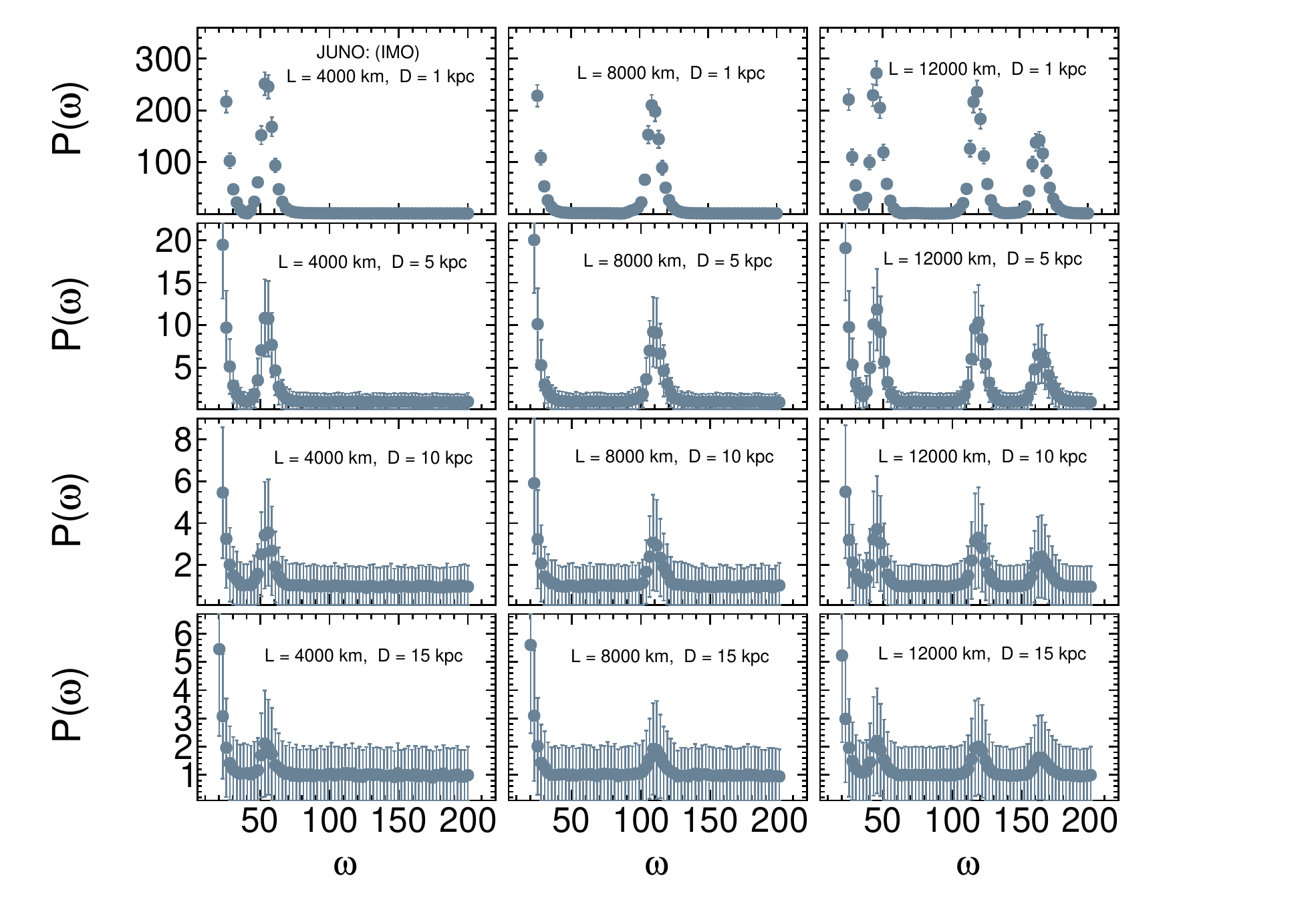}
\vglue -0.4cm
\caption{
CCSN power spectrum at JUNO detector  for several baselines (columns) and distances from the CCSN (rows) for the inverted mass ordering.
We have considered only the scenario S3 (100\% decay) for visualizing the different changes.}
\label{figure:pow_spectrum_misc_IO_JUNO}
\end{center}
\vglue -0.5cm
\end{figure}
\begin{figure}[H]
\begin{center}
\includegraphics[scale=0.7]{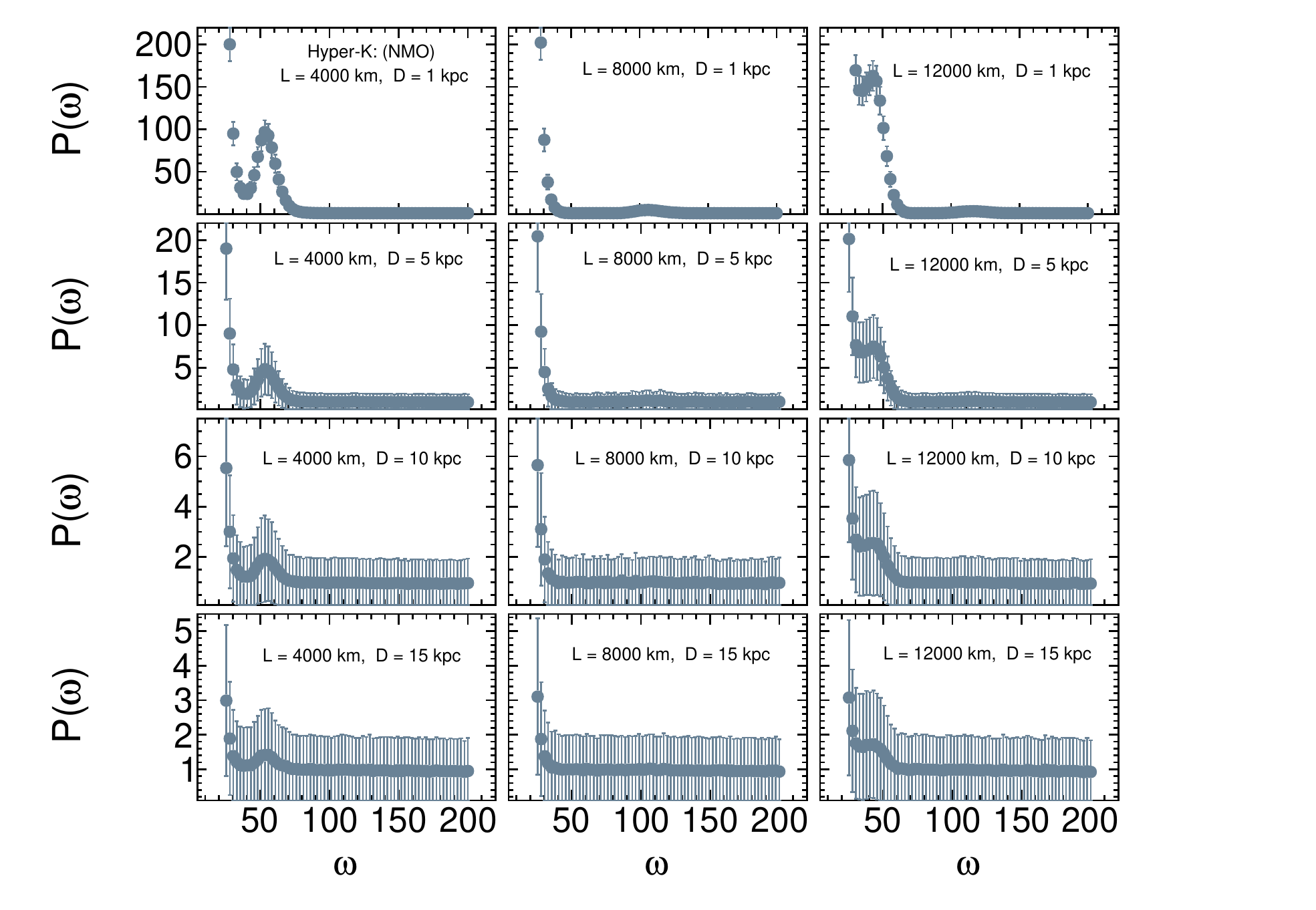}
\vglue -0.4cm
\caption{
CCSN power spectrum at Hyper-K detector for several baselines (columns) and distances from the CCSN (rows) for the normal mass ordering. We have considered only the scenario S3 (100\% decay) for visualizing the different changes. }
\label{figure:pow_spectrum_misc_NO_HK}
\end{center}
\vglue -0.5cm
\end{figure}
\begin{figure}[H]
\begin{center}
\includegraphics[scale=0.7]{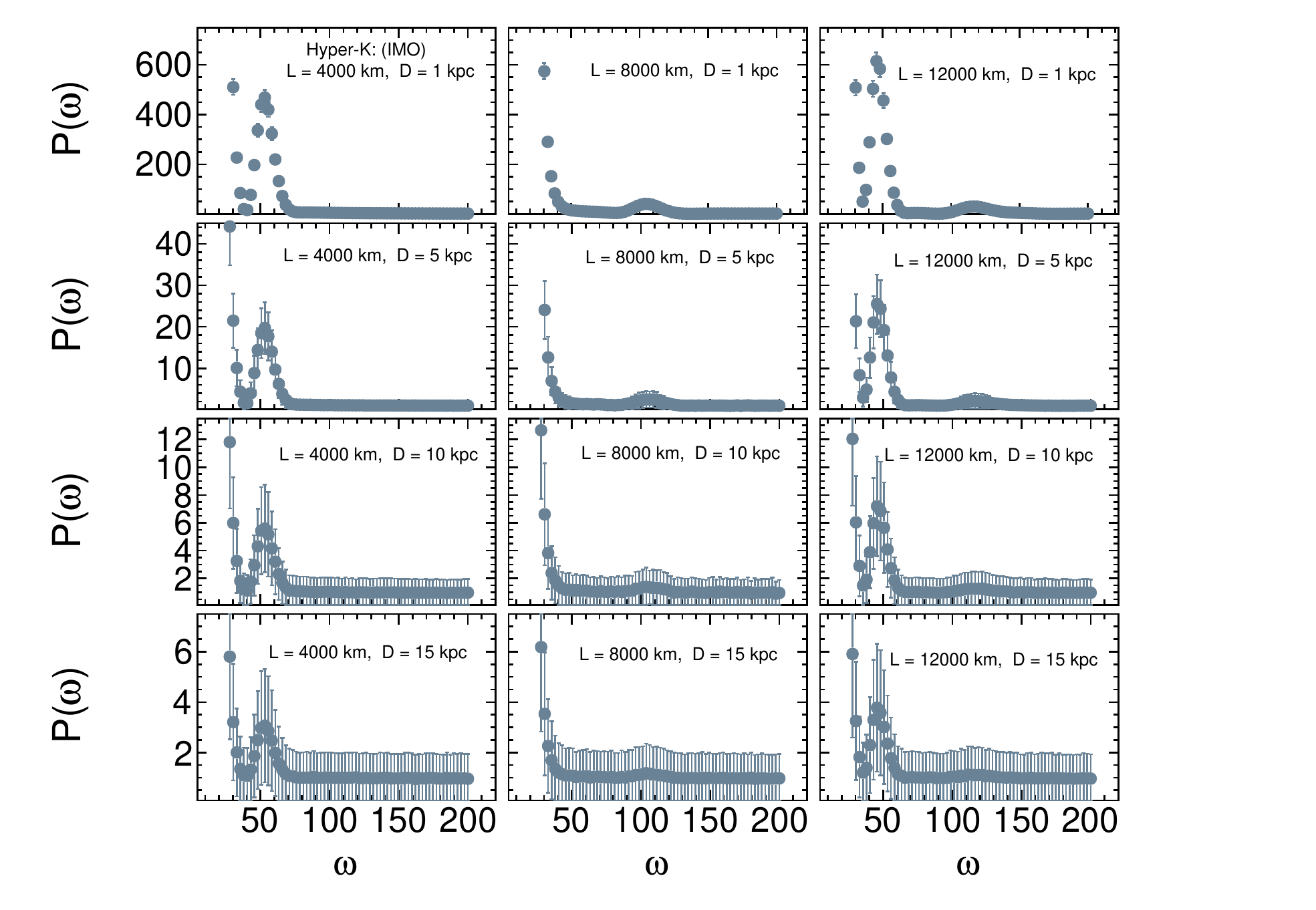}
\vglue -0.4cm
\caption{
CCSN power spectrum at Hyper-K detector for several baselines (columns) and distances from the CCSN (rows) for the inverted mass ordering.  We have considered only the scenario S3 (100\% decay) for visualizing the different changes. }
\label{figure:pow_spectrum_misc_IO_HK}
\end{center}
\vglue -0.5cm
\end{figure}
\begin{figure}[H]
\begin{center}
\includegraphics[scale=0.7]{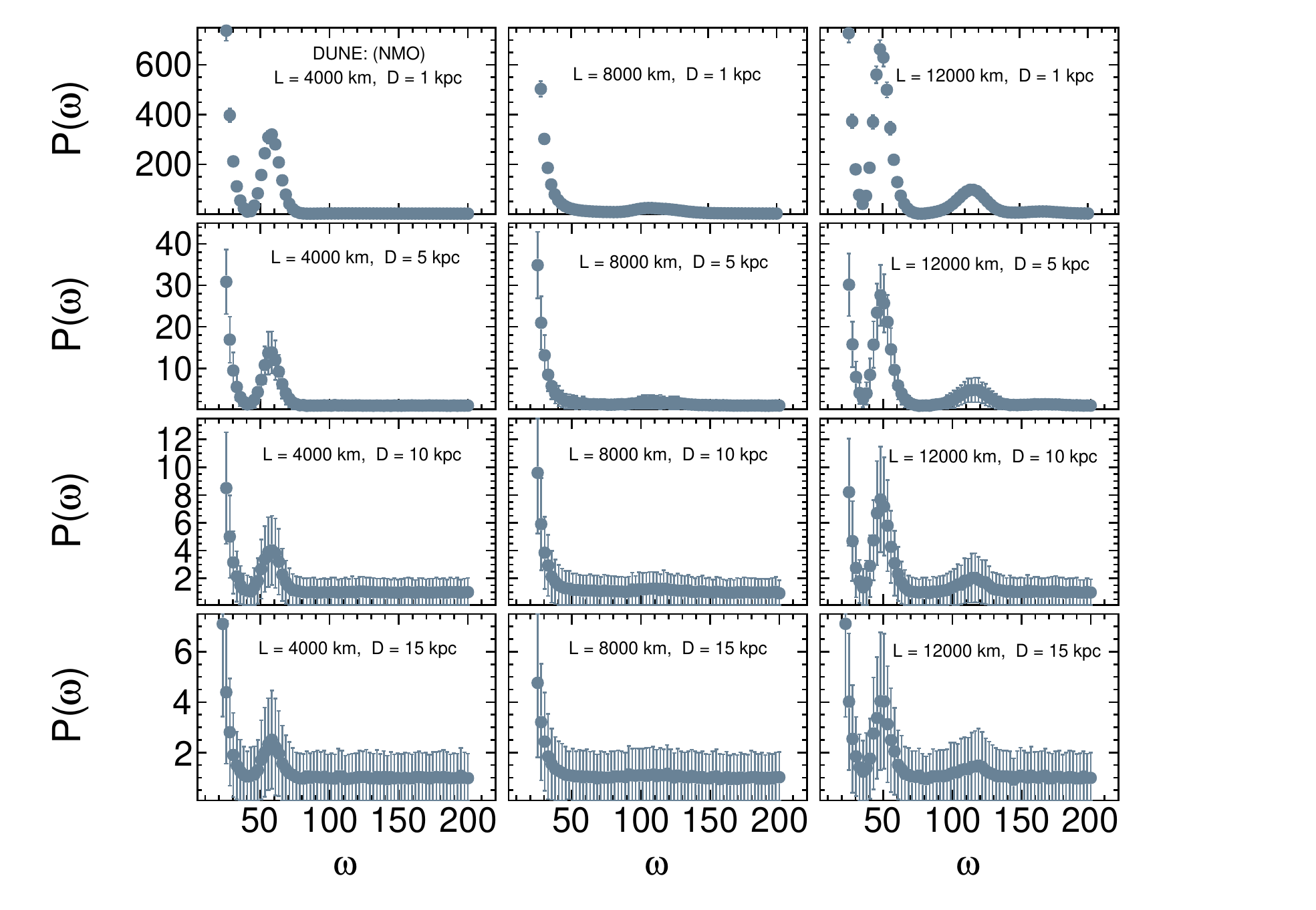}
\vglue -0.4cm
\caption{
CCSN power spectrum at DUNE detector  for several baselines (columns) and distances from the CCSN (rows) for the normal mass ordering. We have considered only the scenario S3 (100\% decay) for visualizing the different changes.}
\label{figure:pow_spectrum_misc_NO_DUNE}
\end{center}
\vglue -0.5cm
\end{figure}
\begin{figure}[H]
\begin{center}
\includegraphics[scale=0.7]{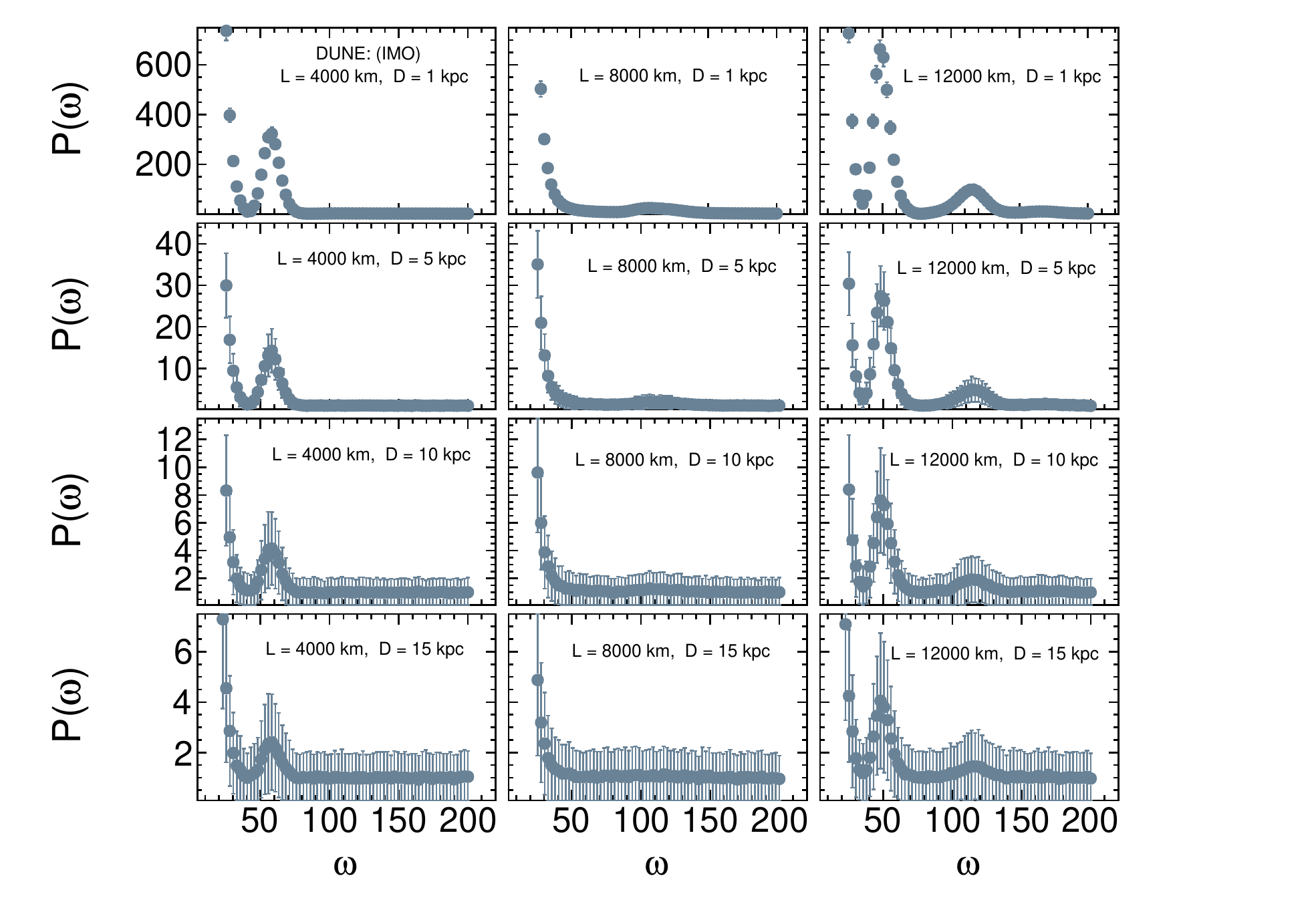}
\vglue -0.4cm
\caption{
CCSN power spectrum at DUNE detector  for several baselines (columns) and distances from the CCSN (rows) for the inverted mass ordering.  We have considered only the scenario S3 (100\% decay) for visualizing the different changes.}
\label{figure:pow_spectrum_misc_IO_DUNE}
\end{center}
\vglue -0.5cm
\end{figure}

\normalem 

\bibliographystyle{JHEP}
\bibliography{references_SN_decay}


\end{document}